\documentclass[12pt]{article}

\usepackage{newtxtext,newtxmath}

\usepackage{graphicx}

\usepackage[mathlines]{lineno}

\usepackage[letterpaper,margin=1in]{geometry}

\renewenvironment{abstract}
	{\quotation}
	{\endquotation}

\date{}

\makeatletter
\renewcommand{\fnum@figure}{\textbf{Figure \thefigure}}
\renewcommand{\fnum@table}{\textbf{Table \thetable}}
\makeatother

\usepackage{scicite}

\usepackage{url}

\def\scititle{

Softness and Hydrodynamic Interactions Regulate Lipoprotein Transport in Crowded Yolk Environments

}
\title{\bfseries \boldmath \scititle}

\author{
    Nimmi Das Anthuparambil$^{1,2\ast}$,
    Michelle Dargasz $^{1}$,
    Sonja Timmermann$^{1}$,\and
    Anita Girelli $^{3}$,
    Sebastian Retzbach$^{4}$,
    Johannes Möller$^{5}$,
    Wonhyuk Jo$^{5}$,\and
    Agha Mohammad Raza$^{1}$,
    Aliaksandr Leonau$^{1,5}$,
    James Wrigley$^{5}$,
    Frederik Unger$^{1,2}$,\and
    Maddalena Bin$^{3}$,
    Prince Prabhu Rajaiah$^{2,6}$, 
    Iason Andronis$^{3}$, 
    William Chèvremont$^{7}$,\and
    Jörg Hallmann$^{5}$,
    Angel Rodriguez-Fernandez$^{5}$,
    Jan-Etienne Pudell$^{5}$, 
    Felix Brausse$^{5}$, \and
    Ulrike Boesenberg$^{5}$, 
    Mohamed Youssef$^{5}$, 
    Roman Shayduk$^{5}$,
    Rustam Rysov$^{5}$,\and
    Anders Madsen$^{5}$,
    Felix Lehmkühler$^{2,8}$,
    Michael Paulus$^{9}$,
    Fajun Zhang$^{4}$,\and
    Fivos Perakis$^{3}$,
    Frank Schreiber$^{4}$, 
    Christian Gutt $^{1\dagger}$\and
 \small$^{1}$Department Physik, Universität Siegen, Walter-Flex-Strasse 3, 57072 Siegen, Germany \and
 \small$^{2}$Deutsches Elektronen-Synchrotron DESY, Notkestr. 85, 22607 Hamburg, Germany\and
 \small$^{3}$Department of Physics, AlbaNova University Center, Stockholm University, S-106 91 Stockholm, Sweden\and
\small$^{4}$Institut für Angewandte Physik, Universität Tübingen, Auf der Morgenstelle 10, 72076 Tübingen, Germany\and
\small$^{5}$European X-Ray Free-Electron Laser Facility, Holzkoppel 4, 22869 Schenefeld, Germany\and
\small$^{6}$Institute for Biochemistry and Molecular Biology, Laboratory for Structural Biology of Infection and \\ \small
Inflammation, University of Hamburg, c/o DESY, 22603, Hamburg, Germany\and
\small$^{7}$ESRF - The European Synchrotron, 71 Avenue des Martyrs, 38042 Grenoble, France\and
\small$^{8}$The Hamburg Centre for Ultrafast Imaging, Luruper Chaussee 149, 22761 Hamburg, Germany\and
\small$^{9}$Fakultät Physik/DELTA, TU Dortmund, 44221 Dortmund, Germany\and
\small$^\ast$Corresponding author. Email: nimmi.das.anthuparambil@desy.de\and
\small$^\dagger$Corresponding author. Email: christian.gutt@uni-siegen.de\and
}

\begin{document} 

\maketitle

\begin{abstract} \bfseries \boldmath
Low-density lipoproteins (LDLs) serve as nutrient reservoirs in egg yolk for embryonic development and as promising drug carriers. Both roles critically depend on their mobility in densely crowded biological environments. Under these crowded conditions, diffusion is hindered by transient confinement within dynamic cages formed by neighboring particles, driven by solvent-mediated hydrodynamic interactions and memory effects—phenomena that have remained challenging to characterize computationally and experimentally. Here, we employ megahertz X-ray photon correlation spectroscopy to directly probe the cage dynamics of LDLs in yolk-plasma across various concentrations. We find that LDLs undergo anomalous diffusion, experiencing $\approx$ 100-fold reduction in self-diffusion at high concentrations compared to dilute solutions. This drastic slowing-down is attributed to a combination of hydrodynamic interactions, direct particle-particle interactions, and the inherent softness of LDL particles. Despite reduced dynamics, yolk-plasma remains as a liquid, yet sluggish, balancing dense packing, structural stability, and fluidity essential for controlled lipid release during embryogenesis.

\end{abstract}

\noindent

Low-density lipoproteins (LDLs) are biological core-shell nanoparticles present in humans, animals, and eggs \cite{gofman1950role,saari1964isolation}. In humans and animals, LDLs play a crucial role in lipid distribution, whereas in egg yolk, they serve as essential lipid reservoirs supporting embryonic development \cite{li2023beyond}. Egg yolk constitutes a naturally crowded medium characterized by high LDL concentrations with volume fractions exceeding 40$\%$. Understanding factors influencing LDL mobility and viscoelasticity in such settings is vital for elucidating yolk stability and effective nutrient delivery to embryos. In human health, hindered LDL diffusion or increased blood viscosity contributes significantly to the formation of atherosclerotic plaques in arterial walls \cite{ference2024ldl, carpenter2022effect}. Furthermore, LDLs have attracted substantial attention as promising drug carriers, exhibiting notable advantages over synthetic delivery systems in navigating complex biological media, which is critical for therapeutic efficiency  \cite{zhu2017biomimetics,wang2021transporting,jutkova2019encapsulation,dai2023low}. Therefore, detailed insights into the mechanisms controlling LDL diffusion under crowded conditions have substantial relevance for biology and medical applications. 

Macromolecular diffusion in densely crowded environments, including the cytoplasm, extracellular matrix, food products, and pharmaceutical formulations, is significantly affected by intermolecular interactions \cite{sun2023effect,nettesheim2020macromolecular,alric2022macromolecular,grimaldo2019dynamics,girelli2021microscopic,anthuparambil2024salt}. Such interactions modulate molecular mobility, creating transient confinement that alter transport properties and molecular function \cite{ramalingam2023macromolecular,bucciarelli2016dramatic,nawrocki2017slow,braun2017crowding}. Crowded conditions typically elevate effective viscosity due to specific electrostatic and non-specific interactions as well as depletion effects \cite{lavalette1999microscopic,zosel2020depletion,heinen2012viscosity}. Moreover, heterogeneous dynamics at elevated concentrations \cite{grimaldo2019protein,beck2022short} often result in deviations from the classical Stokes-Einstein relation, indicating also a possible decoupling of mass and momentum transport \cite{bonn2003stokes,swallen2011self}.

Relevant particle interactions become especially significant at nanometer length and microsecond time scales, where they critically influence diffusion processes and reactions. The characteristic interaction time scale is given by the duration required for a particle, with diffusion constant $D_{0}^\mathrm{d}$, to traverse a distance $R$ comparable to its own size, expressed as $\tau_\mathrm{i}= R^2/6D_{0}^\mathrm{d}$. For LDL particles in egg yolk-plasma this time is approximately $\tau_\mathrm{i} \approx 2.5 \,$ µs, assuming $R=$ 15.5 nm and $D_{0}^\mathrm{d}$ = 15.8 nm$^{2}$/µs. In concentrated solutions, mutual interactions at these time scales lead to memory effects and dynamic caging \cite{Klein1983,NAGALE1996PhyA,BaurPhysRevE1996,weeks2002subdiffusion,weeks2002properties}, significantly influencing particle movement and viscoelasticity. Hydrodynamic interactions further complicate dynamics in dense macromolecular systems \cite{nagele1996dynamics,doster2007microscopic,roosen2011protein,anitaferritin}. These interactions, extending over short and long ranges, yield localized effects at short time scales and correlated collective motion at longer scales \cite{NAGALE1996PhyA}. Despite their importance, the many-body, flow-mediated nature of these hydrodynamic interactions presents ongoing challenges for theoretical modeling and computational simulations due to extensive computational time demands \cite{ando2013dynamic}.

Colloidal theories \cite{brady1988stokesian,holmqvist2010long,TokuyamaPRE1994,tokuyama1995theory} have been widely employed to interpret diffusion in dense systems, predicting behaviors across a wide range of concentrations from dilute to highly concentrated colloidal suspensions, including protein solutions. For instance, spherical proteins such as $\alpha$-crystallin \cite{foffi2014hard,vodnala2018hard} and ferritin \cite{anitaferritin} often follow hard-sphere diffusion predictions, though appropriate scaling might be necessary \cite{roosen2011protein}. However, these simplified models show limitations when applied to macromolecules with complex properties \cite{stradner2020potential,bucciarelli2016dramatic}. Factors such as anisotropy, shape, particle softness, and surface heterogeneity alter interaction potentials and thus may substantially alter diffusion rates \cite{bucciarelli2016dramatic,balbo2013shape,tokuyama2011self,doster2007microscopic}.

Given these limitations, the study of LDL diffusion within its native, densely concentrated egg yolk-plasma environment gains particular importance. LDLs, with their distinct core-shell structure (Fig.~\ref{fig:Experiment_schematic}A) and intricate lipid-protein surface chemistry, exemplify complex biological nanoparticles expected to significantly deviate from idealized hard-sphere predictions. Thus, understanding the factors influencing LDL diffusion is essential both for gaining insights into biological processes involving nutrient release during embryogenesis and enhancing the efficacy of LDL-based drug delivery systems.

Here, we investigate the effects of hydrodynamic interactions and particle softness on LDL diffusion in highly concentrated egg yolk-plasma using megahertz X-ray photon correlation spectroscopy (MHz-XPCS) at the European X-ray Free Electron Laser Facility (EuXFEL) \cite{madsen2021materials}. Our results reveal that LDL diffusion in egg yolk-plasma is strongly influenced by solvent-mediated hydrodynamic interactions, resulting in non-Brownian diffusion \cite{srinivasan2024breaking}. Memory effects stemming from both direct interparticle forces and long-range hydrodynamic interactions at cage-relaxation times lead to a stretched exponential decay in correlation functions. We demonstrate that the $q$-dependence of collective diffusion at interaction time scales is effectively captured by a hydrodynamic function with a scaling prefactor. At short-time scales, this dependence is dominated by direct correlations from the structure factor and hydrodynamic function, while a mean-field prefactor captures the slowing-down effects. The long-time self-diffusion of LDLs in pure yolk-plasma is approximately 100 times slower than in dilute solutions, emphasizing the pronounced influence of crowding. Furthermore, the measured self-diffusion is 3–7 times slower than predictions for hard-spheres, with larger deviations at higher concentrations due to particle softness and hydrodynamic interactions. Pronounced caging effects are observed in the dynamics explaining the microscopic origin of the viscoelasticity of LDL solutions. 

At the highest physiological concentration, the short- and long-time diffusion coefficients differ by a factor of $\approx$ 9, a remarkably large difference, indicative of dynamics approaching a glass-like state. Yet, yolk-plasma remains a viscous liquid. This pronounced separation reflects competing physical requirements: the need for dense, almost glass-like packing to maximize lipid storage, and sufficient elasticity to ensure structural stability. At the same time, molecular mobility must be maintained to prevent LDL coalescence and to enable efficient lipid release during embryogenesis.

\section*{RESULTS}
\begin{figure} 
\centering
\includegraphics[width=0.7 \textwidth]{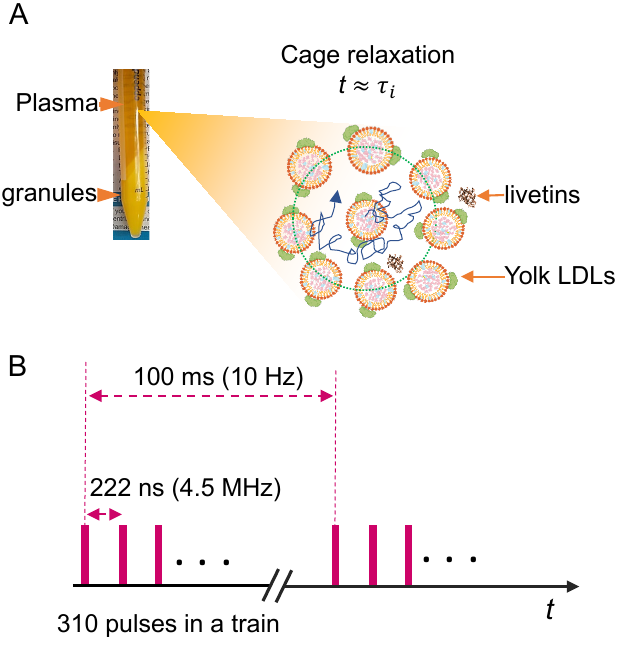} 
\caption{\textbf{EuXFEL time scales access molecular diffusion of yolk LDLs:} (\textbf{A}) The photo on the left shows the two fractions of yolk: yolk-plasma (translucent) and yolk-granules (opaque), separated from egg yolk after centrifugation (see Methods and Materials section). The schematic on the right indicates the cage relaxation of LDL molecules in yolk-plasma. The interaction time $\tau_\mathrm{i}$ of LDL is $\approx$ 2.5 µs. (\textbf{B}). The time structure of European XFEL X-ray pulses used in this experiment. This experiment utilizes X-ray trains with an intra-train pulse frequency of 4.5 MHz and an inter-train repetition rate of 10 Hz. This configuration provides access to an experimental time scale of 0.22--69 µs.}
\label{fig:Experiment_schematic} 
\end{figure}

To investigate LDL diffusion under crowded conditions, we use yolk-LDLs in their native suspension form of yolk-plasma, which was extracted from egg yolk via centrifugation (see Materials and Methods for details). Figure~\ref{fig:Experiment_schematic}A (left side) illustrates the two distinct fractions obtained after centrifugation: the translucent yolk-plasma fraction floating atop the opaque yolk-granule fraction. The yolk-plasma dry matter comprises approximately $\approx$ 85 wt$\%$ yolk-LDL particles and $\approx$ 15 wt$\%$ livetin proteins \cite{anton2013egg}. The LDL particles exhibit a core-shell structure (Fig.~\ref{fig:Experiment_schematic}A) with an average radius of $\approx$ 15.5 nm (see Fig.~\ref{figSI:ff_fit}A). The original LDL concentration in yolk-plasma was $\approx$ 814 mg/ml and was further diluted to two lower concentrations of 668 mg/ml and 547 mg/ml (see Materials and Methods). Corresponding volume fractions were calculated from these weight fractions using the LDL density of 0.98 g/cm$^{3}$ \cite{kuang2018impact}, with resulting values summarized in Table~\ref{tab:1}. 

Samples were loaded into quartz-glass capillaries, sealed with epoxy glue, and probed using MHz-XPCS at the Materials Imaging and Dynamics (MID) instrument \cite{madsen2021materials} at the European X-ray Free Electron Laser facility. The experimental X-ray pulse pattern is illustrated in Fig.~\ref{fig:Experiment_schematic}B. X-ray pulse trains containing 310 pulses with a spacing of 222 ns interacted with the samples, and pulse-resolved scattering patterns were recorded by the Adaptive Gain Integrating Pixel Detector (AGIPD), positioned {7.68 m} from the sample. This configuration provided access to wavevectors in the range $q$ = 0.05–1.1 nm$^{-1}$ and enabled measurements on time scales from 0.22 to 69 µs, covering the interaction time scale ($\tau_\mathrm{i}$ = 2.5 µs) relevant to LDL particle dynamics. To prevent radiation damage, the X-ray intensity on the samples was carefully attenuated, and each sample spot was exposed only once to a single pulse train, maintaining the X-ray dose and dose rate below threshold levels. Further details on dose estimation and beam damage thresholds are provided in the Materials and Methods section and the supplementary information (SI). Statistically reliable data were ensured by averaging scattering patterns and correlation functions collected from thousands of different positions along each capillary. In addition to XPCS measurements, small-angle X-ray scattering (SAXS) measurements were performed at the ID02 beamline at the European Synchrotron Radiation Facility (ESRF), France. The XPCS measurements provided insights into both structural and dynamical characteristics of LDL suspensions, which we present starting with the structural analysis, followed by dynamic properties.

\subsection*{Structural information}

The normalized scattering intensity, $I(q)$, for four concentrations of egg yolk-plasma is shown in  Fig.~\ref{fig:Iq_Sq}A. A pronounced peak at $q \approx 0.23$\, nm$^{-1}$ is observed in the native yolk-plasma ($c=$ 814 mg/ml), reflecting strong interparticle correlations due to the high concentration of LDLs, which account for $\approx$ 85 wt$\% $ of the plasma's dry matter. The scattering profile of the most diluted sample ($c=$ 4 mg/ml) is modeled using a spherical core-shell particle with a core radius of 12 nm and a shell thickness of 3.5 nm, including a generalized Guinier-Porod contribution to capture the scattering contribution of apolipoproteins in the corona. The fit was performed using the SASfit software \cite{bressler2015sasfit}, with full details provided in the SI. As the concentration increases, a change in the size of LDL particles is expected due to their soft, deformable nature and the limited space available, consistent with prior observations of lipid vesicles \cite{quinn2022crowding} and micelles \cite{ogino1988micelle}. As shown in Fig.~\ref{figSI:kratky}, a subtle shift in the second and third minima of $I(q)$ is observed, indicative of structural changes at high concentrations. To estimate the effective form factor at high concentrations, scattering data for $q > 0.65$ nm$^{-1}$ were modeled as described in the SI. Note that the scattering intensity $I(q)$ measured at ESRF and EuXFEL exhibits good agreement, as shown in Fig.~\ref{figSI:saxs_xfel_esrf}. For form factor modeling, the $I(q)$ obtained from ESRF was used due to the extended $q$-range, which provides improved resolution of structural features. Form factor modeling reveals a modest $\approx 3\%$ reduction in LDL size at the highest concentration ($c$ = 814 mg/ml) relative to the dilute case ($c$ = 4 mg/ml).

The structure factor, $S(q)$, is estimated from the experimental $I(q)$ and respective form factor models given in Fig.~\ref{figSI:ff_fit}B-D using Eq.~\ref{eq:S_q}. The estimated $S(q)$ profiles of all concentrations are presented in Fig.~\ref{fig:Iq_Sq}B. The full $S(q)$ profiles across the entire $q$-range are given in Fig.~\ref{figSI:full_Sq}, where $S(q)\to 1$ at high $q$ values illustrates the expected convergence to unity.

\begin{table} 
	\centering
	\caption{\textbf{Details of sample and estimated quantities:} $q_\mathrm{m}$ denotes the $q$-position of the maximum of the structure factor, $S(q_\mathrm{m})$ is the maximum value of the structure factor, $\kappa$ is the isothermal compressibility and $r_\mathrm{cage}$ is the size of the cage deduced from the mean square displacement analysis described in Discussion section.\\}
	\label{tab:1} 	
	\begin{tabular}{|c|c|c|c|c|c|c|c|} 
		\hline
		LDL  &relative amount&relative amount& volume & $q_\mathrm{m}$ & $S(q_\mathrm{m})$ & $\kappa$ & $r_\mathrm{cage}$ \\
        concen- & of yolk-plasma & of salt-buffer$^{*}$ & fraction,  & & & & \\
         tration& in the solution & added & $\phi$ & & & & \\
        \hline
		[mg/ml] & [wt$\%$] & [wt$\%$] & & [nm$^{-1}$] & &[kPa$^{-1}$]& nm\\
		\hline
		814 & 100 & 0 &  0.43  & 0.235 & 2.79 & 2 & 4.4$\pm$0.2\\ 
		668 & 90 & 10 &  0.39 & 0.226 & 2.48 & 3.6 & 6$\pm$0.2\\
		547 & 80 & 20 &  0.34  & 0.217 & 2.16 & 5.1 & 8.2$\pm$0.3\\

		\hline
  \multicolumn{4}{l}{\small $^{*}$ 170 mM NaCl solution.} \\
	\end{tabular}
\end{table}

The structure factor peak position, $q_\mathrm{m}$, shifts to lower $q$-values with decreasing LDL concentration, indicating an increase in the interparticle correlation distance $d_\mathrm{m} = 2\pi/q_\mathrm{m}$. As expected, the peak height of the structure factor, $S(q_\mathrm{m})$, increases with increasing LDL concentration (see Table~\ref{tab:1}). A detailed analysis of $S(q)$ at low $q$-values reveals a relatively high asymptotic limit, $S(q \to 0) = S(0) > 0.25$, for all LDL solutions with $\phi > 0.3$ (Table~\ref{tab:1}). This contrasts with typical hard-sphere suspensions, which exhibit $S(0) < 0.1$ at similar volume fractions. The larger $S(0)$ observed for LDL suspensions indicates increased isothermal compressibility ($\kappa$), as $S(0)$ is directly proportional to $\kappa$ (see Eq.~\ref{eq:I_C}). To confirm this finding, we estimated the compressibility of LDL solutions from experimental $S(0)$ values and compared these results to theoretical predictions for hard-spheres at equivalent volume fractions (see inset of Fig.~\ref{fig:Iq_Sq}B). Remarkably, LDL suspensions exhibit approximately nine-fold higher compressibility than hard-sphere suspensions at comparable volume fractions, underscoring the inherent softness of LDL particles.

\subsection*{Wave-vector dependent dynamics}
\begin{figure} 
	\centering
	\includegraphics[width=0.95 \textwidth]{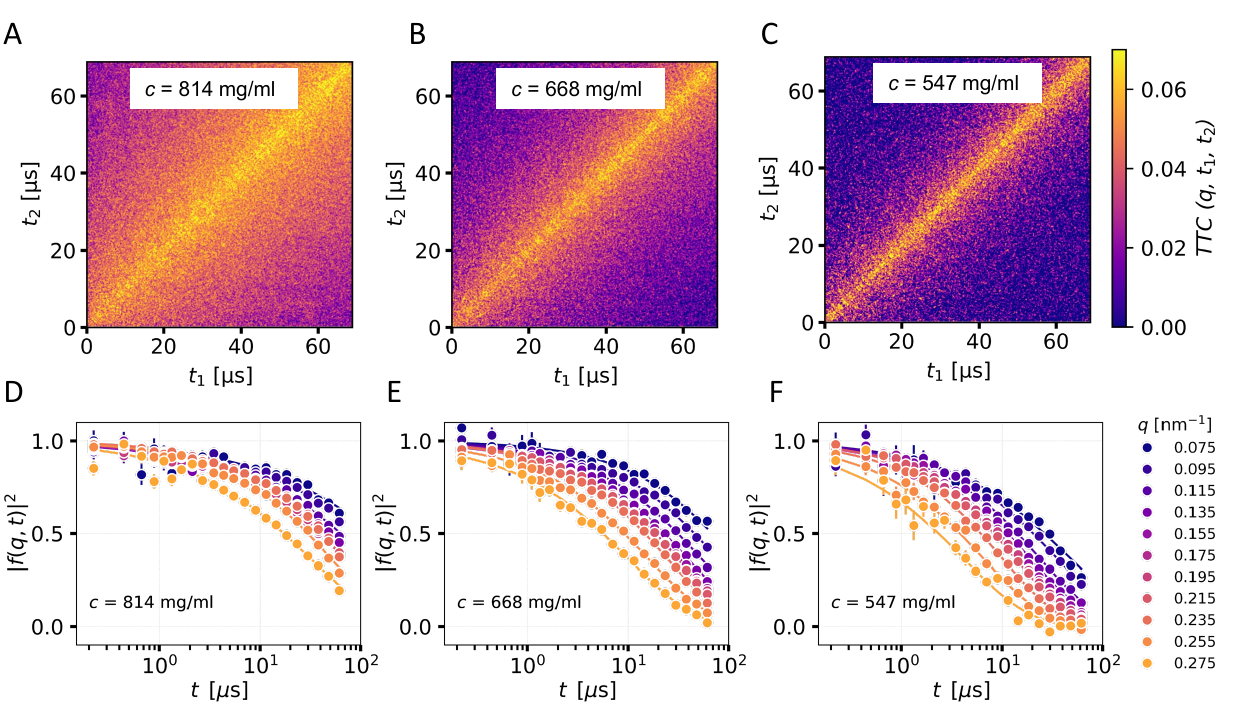} 
	\caption{\textbf{MHz-XPCS data of yolk-plasma solutions at different concentrations: } (\textbf{A-C}) Two-time correlation functions of yolk-plasma at different concentrations. The $TTC$s are extracted at $q = 0.225 \, \text{nm}^{-1}$. (\textbf{D-F}) $|f (q, t )|^2$ extracted from the $TTC$s at different $q$ values. The solid lines indicate the fits using Eq.~\ref{eq:g2}.
}
	\label{fig:TTC_g2} 
\end{figure}

To estimate the relaxation times of the LDL particles, we calculated the two-time correlation function \cite{perakis2020towards} defined as
\begin{equation}
    TTC (q,t_{1},t_{2}) = \frac{\langle I_{p}(q,t_{1}) I_{p}(q,t_{2})\rangle}{\langle I_{p}(q,t_{1}) \rangle \langle I_{p}(q,t_{2})\rangle},
\label{eq:TTC}
\end{equation}
\noindent where $I_{p}$ denotes the scattering intensity at pixel $p$, and the angular brackets $\langle ... \rangle$ represent an average over pixels with the same $q$ value. The variables $t_{1}$ and $t_{2}$ are different experimental time points. Examples of the extracted $TTC$ maps at $q = 0.225\, \mathrm{nm}^{-1}$ are shown in Fig.~\ref{fig:TTC_g2}A-C for all three LDL concentrations studied. The width of the bright diagonal (yellow) region in the maps is inversely related to the relaxation rate of the particle dynamics. The narrowest diagonal for the least crowded sample (Fig.~\ref{fig:TTC_g2}C) indicates rapid particle motion, while broader diagonals observed for higher concentrations (Fig.~\ref{fig:TTC_g2}A-B) clearly demonstrate a substantial slowing of particle dynamics with increased crowding.

\begin{figure} 
	\centering
	\includegraphics[width=0.95 \textwidth]{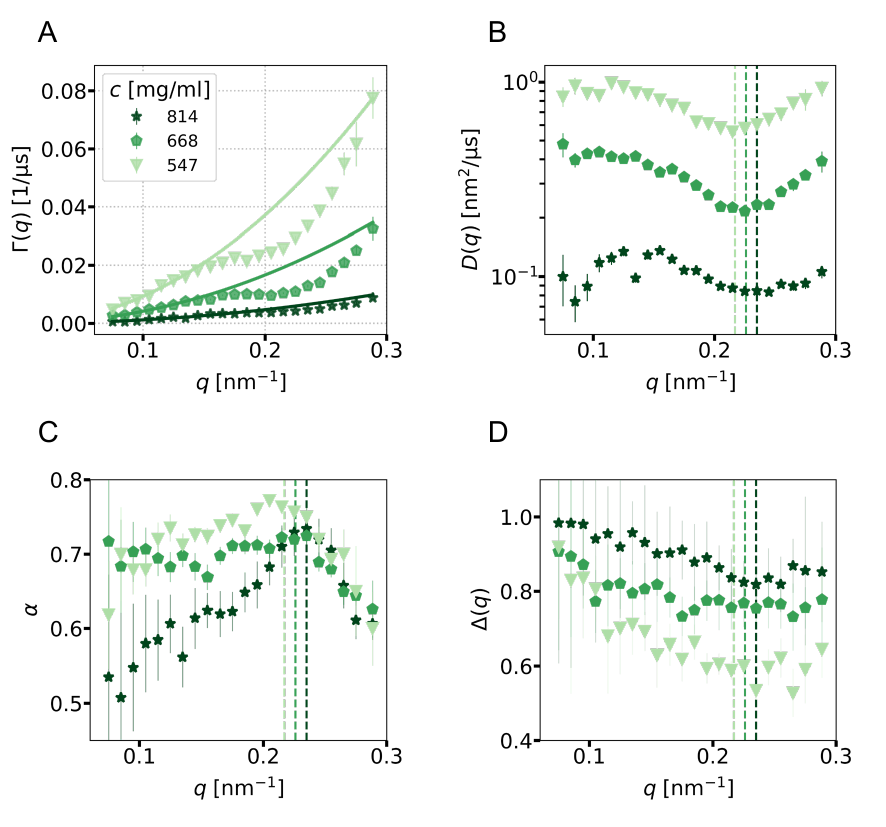} 
	\caption{\textbf{Length-scale dependent dynamics of LDLs:} (\textbf{A}) Wavevector dependent relaxation rate of yolk-plasma at different concentrations. The solid line is the fit using $\Gamma(q) = \tilde{D} \; q^{2}$, performed at the first 6 data points ($q$-range of 0.075--0.125 $\text{nm}^{-1}$), where $\tilde{D}$ is a constant. The strong deviation of experimental data points at $q > 0.15$ $\text{nm}^{-1}$ from the fit indicates the de Gennes narrowing. (\textbf{B}) The wave vector-dependent diffusion coefficient extracted from the $\Gamma (q)$ using the relation $D(q) = \Gamma (q) / q^{2}$. (\textbf{C}) The wavevector-dependent KWW exponent and (\textbf{D}) reduced memory function (or non-exponentiality function) of all samples. The dashed vertical lines in \textbf{B}-\textbf{D} indicate the peak position of the experimental structure factor ($q_\mathrm{m}$).
}
	\label{fig:Gammaq_Dq} 
\end{figure}

From the $TTC$ maps, we obtained intensity autocorrelation functions $g_{2}(q,t)$ by horizontal averaging \cite{anthuparambil2023exploring}. These were modeled using:
\begin{equation}
g_{2} (q,t) = 1+\beta (q)|f(q,t)|^2 = 1 + \beta(q) \; \vert \; \text{exp}[-(t \,\tilde{\Gamma}(q) )^{\alpha(q)}] \; \vert^{2},
\label{eq:g2}
\end{equation}

\noindent 
where $f(q,t)$ is the intermediate scattering function, $\tilde{\Gamma}(q)$ is the $q$-dependent relaxation rate, $\beta(q)$ is the speckle contrast, and $\alpha(q)$ is the Kohlrausch-Williams-Watts (KWW) exponent \cite{williams1970non}, describing the shape of the correlation function.\\
Representative intermediate scattering functions and corresponding fits are shown in Fig.~\ref{fig:TTC_g2}D-F. We observe a pronounced stretching of the correlation functions with KWW exponents varying from 0.5 to 0.8 (Fig.~\ref{fig:Gammaq_Dq}C), indicating greater stretching (lower $\alpha$) at higher concentrations. From the stretched correlation functions we extract mean relaxation rates via $\Gamma(q) = \tilde{\Gamma}(q) \,\alpha(q) \, / \Gamma_\mathrm{f}(1/\alpha(q))$, where  $\Gamma_\mathrm{f}$ denotes the Gamma function \cite{arbe2009characterization,ruta2020wave}. The resulting $\Gamma(q)$ values are depicted in Fig.~\ref{fig:Gammaq_Dq}A, and we observe a decrease of $\Gamma(q)$ with increasing concentration, indicating a slowdown in the particle dynamics as expected. In all samples, we find a $\Gamma (q) \sim q^2$ (solid lines in Fig.~\ref{fig:Gammaq_Dq}A) behavior at low $q$ values, which is typical of Brownian dynamics. However, $\Gamma (q)$ significantly deviates from the Brownian behavior at $q$ values near the structure factor peak at $q_{m}$. Such modulations in $\Gamma(q)$ around the $q_{m}$ are typical signatures of collective relaxation phenomena (de Gennes narrowing: $\Gamma(q) \propto 1/S(q)$\;) observed in dense fluids, protein solutions, colloids, and glasses \cite{braun2017crowding,hong2014gennes,nygaard2016anisotropic,ruta2020wave,caronna2008dynamics}. \\

Next, we extracted the $q$-dependent collective diffusion coefficient $D(q)$ using the relation $D(q) = \Gamma(q)/q^2$ (Fig.~\ref{fig:Gammaq_Dq}B). The resulting $D(q)$ exhibits a minimum at $q = q_\mathrm{m}$, indicating that density fluctuations with wavelengths corresponding to the typical cage structures are slowed down relative to other modes. This behavior suggests a coupling to the local structure of the suspension. Given that the interaction time of LDLs ($\approx 2.5$ µs) falls within the experimental time window (0.22--69 µs), we are indeed probing the cage relaxation dynamics of LDLs (Fig.~\ref{fig:Experiment_schematic}A). At the peak of the static structure factor, these modes correspond to nearest-neighbor distances, thus providing access to fluctuations of the dynamic cages formed by neighboring particles.

We also examine the wavevector-dependence of the KWW-exponent $\alpha$ as shown in Fig.~\ref{fig:Gammaq_Dq}C. For $q < q_\mathrm{m}$, $\alpha$ shows a weak decreasing trend with increasing concentration. Interestingly, the $\alpha$ values exhibit an $S(q)$-like modulation near the structure factor peak ($q > 0.15 \;\mathrm{nm}^{-1}$), with the effect becoming more pronounced at the highest LDL concentration. Similar behavior has been observed in metallic glasses near the $S(q)$ peak \cite{ruta2020wave,neuber2022disentangling,luo2022q}. We attribute this modulation to pronounced memory effects in the LDL dynamics on time scales $t \gtrsim \tau_\mathrm{i}$ \cite{Klein1983,NAGALE1996PhyA,BaurPhysRevE1996}, leading to a stretched exponential decay of the intermediate scattering function $f(q,t)$ and consequently to $\alpha < 1$.

This interpretation can be understood by considering Eq.~\ref{eq:memory_eq}, which relates $f(q,t)$ to the collective memory function. In the short-time regime ($t \ll \tau_\mathrm{i}$), the memory term (i.e., the integral involving $M(q,t)$) is negligible, resulting in an exponential decay of $f(q,t)$ \cite{NAGALE1996PhyA}. However, at longer times, the memory term becomes significant, capturing the influence of past particle configurations on current dynamics, a hallmark of caging. The degree of stretching in $f(q,t)$, and thus the deviation of $\alpha$ from unity, reflects the strength of these memory effects. A quantitative measure of this behavior is given by the reduced memory function (or non-exponentiality function) \cite{Klein1983,NAGALE1996PhyA,BaurPhysRevE1996}

\begin{equation}
    \Delta (q) = 1- \frac{\Gamma(q)}{\Gamma^{s}(q)},
\end{equation}
\noindent where $\Gamma^{s}(q)$ denotes the relaxation rate in the short-time limit $t \ll \tau_\mathrm{i}$ during which $f (q,t)$ decays exponentially \cite{NAGALE1996PhyA}. To estimate $\Gamma^{s}(q)$, Eq.~\ref{eq:g2} was fitted to the data with $\alpha$ = 1 for $t<\tau_\mathrm{i}$ (see SI for details on the extraction procedure). Weak and strong memory effects are characterized by  $\Delta(q)= 0$ and $\Delta(q) = 1$, respectively. Fig.~\ref{fig:Gammaq_Dq}D shows the experimentally extracted reduced memory function for LDLs. Overall, $\Delta(q)$ increases with LDL concentration, indicating enhanced memory effects in denser suspensions. Additionally, in all samples, $\Delta(q)$ decreases with increasing $q$, and at the highest concentration, $\Delta(q)$ exhibits a local minimum near the $S(q)$ peak position. This feature is consistent with previous experimental observations and theoretical predictions based on mode-coupling approximations for moderately polydisperse colloidal suspensions \cite{BaurPhysRevE1996,taylor1985memory,hartl1992structure}.

\section*{DISCUSSION}
In the absence of hydrodynamic interactions, the wave-vector dependence of $D(q)$ would entirely result from the structural correlations, following the relation $D(q) = D_0/S(q)$\cite{hess1983generalized}. However, comparison of $D_{0}/D(q)$ and $S(q)$ in Fig.~\ref{figSI:Sq_D0_Dq} reveals significant discrepancies in their $q$-dependence, indicating the presence of strong hydrodynamic interactions. 

\begin{figure} 
	\centering
	\includegraphics[width=0.95 \textwidth]{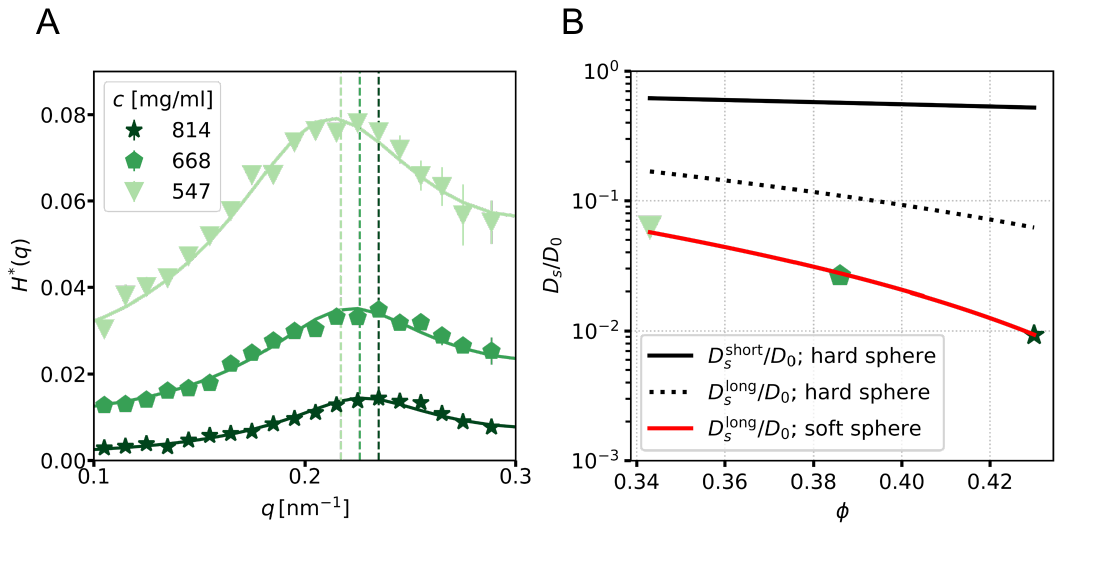} 
	\caption{\textbf{Hydrodynamic interactions and normalized self-diffusion coefficient:}  (\textbf{A}) Experimental hydrodynamic functions (solid data points) along with model fits (solid line) as described in the text. Dashed vertical lines indicate the peak position of the experimental structure factor ($q_\mathrm{m}$). (\textbf{B}) The self-diffusion coefficient of LDLs as a function of the volume fraction of LDLs: solid black curve: Eq. \ref{eq:Ds_D0_tokuyama}, black dotted curve: Eq. \ref{eq:DL_D0_tokuyama} with $\epsilon=1$, and solid red curve: Eq. \ref{eq:DL_D0_tokuyama} with $\epsilon=2$.
}
	\label{fig:Hq_Dsq} 
\end{figure}

Given the lack of theoretical expressions for $D(q)$ valid on the interaction time scale, we introduce an effective hydrodynamic function $H^{*}(q)$ to interpret our results. This function mirrors the familiar form for short-time dynamics  \cite{nagele1996dynamics,dallari2021microsecond}:
\begin{equation}
    D(q) =  \frac{D_{o} H^{*}(q)}{S(q)},
\label{eq:Dq_Hq}
\end{equation}
\noindent where $H^*(q)$ captures hydrodynamic interactions on the intermediate time scales relevant to our experiment. It is related to the short-time hydrodynamic function by $H^*(q) = \frac{D_s^\mathrm{long}}{D_s^\mathrm{short}} H(q)$, where $D_s^\mathrm{long}$ and $D_s^\mathrm{short}$ are the long- and short-time self-diffusion coefficients, respectively. Rescaled hydrodynamic functions of this kind have previously been employed to model data from hemoglobin solutions \cite{doster2007microscopic} and colloidal suspensions \cite{maier2024structure,abade2010short}. This approach enables us to extract $H^*(q)$ from experimental data using Eq.~\ref{eq:Dq_Hq} (Fig.~\ref{fig:Hq_Dsq}A). We find that the peak positions of $H^*(q)$ closely align with those of $S(q)$, as indicated by the vertical dashed lines in Fig.~\ref{fig:Hq_Dsq}A. Moreover, the $q$-dependence of $H^*(q)$ is well captured by the $\delta\gamma$-formalism of Beenakker and Mazur \cite{beenakker1983diffusion,beenakker1984diffusion} (see Materials and Methods and Fig.~\ref{figSI:SI_Hq_fits}). Interestingly, this implies that the $q$-dependence of the collective diffusion coefficient on intermediate time scales—where interparticle interactions dominate—is similar to that observed at short times, where hydrodynamic interactions are dominant. Comparable scaling behavior has been reported for colloids within restricted $q$-intervals \cite{doster2007microscopic}, and more recently by us in dense ferritin solutions \cite{anitaferritin}.
\\
The long wave-vector limit of the short-time hydrodynamic function $H(q\to\infty)$, yields the normalized self-diffusion coefficient $D_\mathrm{s}^\mathrm{short}/{D_{0}}$ \cite{nagele1996dynamics}. Similarly, for the effective hydrodynamic function $H^*(q)$,
the long-$q$ limit gives $H^*(q\to\infty) = D_\mathrm{s}^\mathrm{long}/D_0$. We extract this quantity from our experimental data and plot $D_\mathrm{s}^\mathrm{long}/D_0$ as a function of volume fraction in Fig.~\ref{fig:Hq_Dsq}B.

In our recent study on ferritin \cite{anitaferritin}, the long-time self-diffusion followed the predictions of the hard-sphere-based model by Medina-Noyola \cite{medina1988long} and van Blaaderen et al. \cite{van1992long}, in which direct and hydrodynamic interactions are decoupled. In this framework, hydrodynamic interactions influence only the short-time self-diffusion, while direct interactions are described through the contact value of the pair correlation function.

However, this model fails to account for the significantly reduced values of $D_\mathrm{s}^\mathrm{long}/D_0$ observed for LDL particles (see Fig.~\ref{figSI:medina}). This discrepancy arises from neglecting particle softness and long-range hydrodynamic interactions. These effects are incorporated in the theoretical treatment by Tokuyama et al. \cite{tokuyama1995theory,tokuyama2011self}, which provides an expression for the volume-fraction dependence of $D_\mathrm{s}^\mathrm{long}/D_0$ that includes both short- and long-range hydrodynamic interactions:

\begin{equation}
\begin{gathered}
\frac{D_\mathrm{s}^\mathrm{long}}{D_{0}} = \frac{1-(9\phi/32)}{1+L(\phi)+\epsilon K(\phi)},\\
K(\phi) = \frac{(\phi/\phi_{0})}{(1-\phi/\phi_{0})^{2}},\\
\epsilon = \int_{1}^{\infty} \mathrm{d}\hat{r} \hat{r}^{3} \biggl(  \frac{\partial \hat{U}(r)}{\partial \hat{r}} \biggr) \\
\end{gathered}
\label{eq:DL_D0_tokuyama}
\end{equation}
\noindent where $U(r)$ is the interparticle interaction potential. For a hard-sphere potential, $\epsilon = 1$, while for a soft repulsive interaction, such as an inverse power-law potential $U(r) = k_\mathrm{B}T \left(\frac{\sigma}{r}\right)^6$, $\epsilon = 2$, with $\sigma$ denoting the effective particle diameter. The expression for $L(\phi)$ is provided in the Materials and Methods section. The parameter $\phi_0$ represents a singular volume fraction which is determined by the many-body long-range hydrodynamic interactions between particles. The numerator term, $(9\phi/32)$, arises from the coupling between direct interactions and short-range hydrodynamic interactions. The function $K(\phi)$, which scales strongly with volume fraction, encapsulates many-body correlation effects due to long-range hydrodynamic interactions and is the primary contributor to the significant reduction in $D_\mathrm{s}^\mathrm{long}$ (see Fig.~\ref{figSI:SI_Tokuyama_DL_Do}) \cite{TokuyamaPRE1994}.

The prediction of $D^\mathrm{long}_{s}/D_{0}$ for this model, assuming hard-sphere interactions only (dotted line in Fig.~\ref{fig:Hq_Dsq}B), overestimates the diffusion constants also by a factor of 3--7. In contrast, $D^\mathrm{long}_{s}/D_{0}$ for soft-spheres aligns with the data (red line in Fig.~\ref{fig:Hq_Dsq}B), highlighting the critical role of particle softness in limiting the self-diffusion coefficients of LDL particles in egg yolk-plasma. Furthermore, the critical volume fraction obtained from the fit is $\phi_{0} = 0.5$, which is considerably lower than the predictions for hard-spheres ($\phi_{0} = 0.5718$) \cite{tokuyama1995theory}. Interestingly, similarly low critical volume fractions have been reported for other biomolecules, such as bovine serum albumin (BSA) ($\phi_{0} = 0.49$) \cite{begam2020packing} and lysozyme ($\phi_{0} = 0.34$) \cite{bergman2019experimental}. Notably, the diffusion coefficients of BSA were also found to match predictions for soft spheres \cite{balbo2013shape}, further illustrating the impact of particle softness on both diffusion coefficients and critical volume fractions. We note that this observation is in good agreement with the finding that the isothermal compressibility of the LDL particle solutions is $\approx$ 9 times higher than that of hard-spheres (inset of Fig.~\ref{fig:Iq_Sq}B), indicating the increased particle softness and hence supports the findings here.

We examine the validity of the Stokes–Einstein relation by comparing the effective viscosity ($k_B T/6\pi R D_\mathrm{s}^\mathrm{long}$) estimated from XPCS with viscosity values obtained from rheology measurements in Fig.~\ref{figSI:viscosity}. The shear-dependent viscosity is consistently slightly higher than that estimated from XPCS utilizing $D_\mathrm{s}^\mathrm{long}$ values. This behavior has previously also been observed in colloidal systems \cite{bonn2003stokes} and small organic compounds \cite{swallen2011self} at high concentrations near the glass transition. This decoupling of momentum and mass transport, and hence deviations from the Stokes-Einstein relation, is a hallmark of dynamics approaching the glass transition \cite{bonn2003stokes}.

Additional insight into the nature of the stretched correlation function can be obtained by evaluating the mean-square displacement (MSD) from the intermediate scattering function. For this, we first extract the collective width function, $w(q,t)$, from $f(q,t)$ using the expression \cite{banchio2018short,martinez2011dynamics},
\begin{equation}
    w (q,t) = -\,\frac{1}{ q^{2}} \; \mathrm{ln} [f(q,t)].
\label{eq:width_fn}
\end{equation}
\noindent 
The incoherent MSD can then be estimated from $w(q,t)$ for $qR \ge 2.5$ using the relation \cite{segre1996scaling,segre1997dynamics,banchio2018short}
\begin{equation}
    \mathrm{MSD}\, (t) = 6\,w(q,t) \frac{D_\mathrm{s}^\mathrm{long}}{D(q)},
\label{eq:msd}
\end{equation}
\noindent which is an empirical expression from experiments on colloidal hard-sphere suspensions \cite{segre1996scaling,segre1997dynamics}. The validity of this expression for yolk-plasma solutions is confirmed by the collapse of all MSD curves for $qR \ge 2.5$ as shown in Fig. \ref{figSI:MSD_diffq}. Thus, averaging all MSD values displayed in Fig.~\ref{figSI:MSD_diffq} yields a single representative MSD curve for each sample, as shown in Fig. \ref{fig:MSD}A. 

In the short- and long-time regimes, a diffusive behavior (MSD $\sim t$) is expected, denoted by the blue dash-dotted and dashed lines (Fig.~\ref{fig:MSD}A). However, at time scales of $\tau_\mathrm{i}$, the MSD displays a sub-diffusive regime with MSD $\sim t^{\zeta}$ ($\zeta < 1$). In other words, we observe a transition from short-time to long-time diffusive behavior on the experimentally investigated time scales. This non-linear behavior of the MSD is in good agreement with the pronounced stretching of the correlation functions observed (Fig. \ref{fig:Gammaq_Dq}). 

As the concentration increases, the MSD decreases, indicating the expected slowdown associated with crowding. Furthermore, the distinction between short- and long-time regimes becomes increasingly pronounced at higher concentrations. To quantify this effect, the MSD curves are modeled using Eq.~\ref{eq:msd_fit}, and the corresponding $D_\mathrm{s}^\mathrm{short}/D_\mathrm{s}^\mathrm{long}$ values are extracted (inset of Fig.~\ref{fig:MSD}B). Evidently, $D_\mathrm{s}^\mathrm{short}/D_\mathrm{s}^\mathrm{long}$ increases significantly with rising volume fraction, mirroring the trend observed in hard-sphere colloidal systems \cite{holmqvist2010long}, though with comparatively smaller values.

At the highest concentration, the diffusion coefficient differs by a factor of $\approx9$. This value is similar to that seen in monodisperse colloids near freezing \cite{lowen1993dynamical,holmqvist2010long}, which suggests the LDLs are packed very densely. This dense packing serves two primary biological functions: it maximizes the storage of vital lipids for the developing embryo and helps maintain the structural integrity of the yolk. Despite the marked reduction in diffusion rates, the yolk remains in a liquid state (Fig.~\ref{figSI:moduli}B), indicating a \textit{sluggish liquid} phase. This condition is potentially optimal for biological purposes—stable enough to preserve lipid storage while remaining fluid enough for nutrient access by the embryo when required.

Further, we estimate the viscoelastic properties of yolk-plasma samples from the time-dependent MSD fit, following the formalism presented in ref. \cite{mason2000estimating}. The resulting estimates for the storage and loss moduli are shown in Fig.~\ref{figSI:moduli}. The loss modulus exceeds the storage modulus, confirming the liquid-like nature of the samples. A comparison of the theoretical prediction for the storage modulus for hard-spheres (Eq.~\ref{eqSI:storage_HS}) with the plateau value of the storage modulus of LDL particles shown in Fig.~\ref{figSI:moduli}C reveals a clear discrepancy with the hard-sphere model significantly underestimating the measured storage modulus. This is likely attributed to the enhanced elasticity of the cages formed by the strongly localized LDL particles at higher concentrations, indicating stronger interparticle interactions than expected for hard-sphere suspensions. The elevated storage modulus also suggests a pronounced resistance to deformation of the local cage structure, contributing to the overall stability of egg yolk-plasma.

More information on the anomalous nature of diffusion in the intermediate time scales can be obtained by extracting the anomalous exponent $\zeta$ using
\begin{equation}
    \zeta = \frac{\mathrm{d \,ln( MSD}(t) ) }{\mathrm{d \,ln}(t)}
    \label{eq:zeta}
\end{equation}
\noindent The $\zeta$-values estimated from the MSD fit for all samples are displayed in Fig. \ref{fig:MSD}B. Within the experimentally accessible time scales (denoted by the vertical dashed lines), $\zeta$ remains consistently below 1, confirming the sub-diffusive nature of the particle dynamics during caging. As the volume fraction increases, $\zeta$ decreases, indicating a stronger degree of anomalous behavior due to crowding and interactions. 

The strength of this caging effect is quantified by the minimum $\zeta$ value observed within the measurement time window, as shown in the inset of Fig. \ref{fig:MSD}B. As expected, the caging strength increases with volume fraction, highlighting the impact of crowding on the motion of soft LDL particles. The cage size, defined as $r_\mathrm{cage} \approx \sqrt{\text{MSD}(t = t_\mathrm{cage})}$, is estimated from the MSD at the cage rearrangement time $t_\mathrm{cage}$—the point where the MSD starts to rise from its plateau \cite{weeks2002properties}. This plateau corresponds to the minimum $\zeta$ values, making it a good reference for determining $r_\mathrm{cage}$ (see Table 1). As the volume fraction increases, $r_\mathrm{cage}/R$ follows a trend similar to that observed in hard-sphere colloids but with notably smaller values \cite{weeks2002properties}. This suggests that LDL particles achieve comparable $r_\mathrm{cage}/R$ ratios to hard-spheres, albeit at significantly lower volume fractions.

\section*{SUMMARY}

In summary, we investigated the diffusion of yolk LDLs in concentrated solutions at molecular length and interaction time scales, employing MHz-XPCS at EuXFEL. Our findings reveal that LDLs exhibit non-Brownian dynamics in the crowded environment of egg yolk-plasma. The stretched exponential decay observed in the autocorrelation functions suggests the presence of memory effects, originating from both direct particle interactions and long-range hydrodynamic interactions. The long-time self-diffusion coefficients of LDLs, estimated from the hydrodynamic function at three different concentrations, were found to be 3--7 times lower than the predictions for hard-sphere suspensions. This highlights the limitations of the hard-sphere colloidal model for describing the dynamics of soft biomolecules. Instead, the self-diffusion of LDLs in crowded conditions aligns well with the theoretical expectations for soft spheres, where long-range hydrodynamic interactions play a significant role. Additionally, in pure yolk-plasma, we observe a pronounced diffusion ratio ($D_\mathrm{s}^\mathrm{short}/D_\mathrm{s}^\mathrm{long} \approx 9$) alongside a high storage modulus, indicative of dense LDL packing and structural stability. We anticipate that this dense arrangement optimizes lipid storage while maintaining a \textit{sluggish liquid} state that supports controlled nutrient release to the developing avian embryo. Overall, our results demonstrate that seemingly common biological systems, such as LDL in egg yolk, exhibit complex diffusive dynamics driven by the interplay of hydrodynamic interactions and soft interparticle potentials.

\section*{MATERIALS AND METHODS}

\subsection*{Sample preparation}

The egg yolk used in this study was obtained from a hen egg purchased at a supermarket. The yolk-plasma samples used for the experiments are prepared in five steps. Step 1: The yolk was separated from the egg white using a steel strainer and then washed in Milli-Q water to eliminate any excess albumen on the yolk. Step 2: The cleaned yolk was placed on a filter paper to remove any excess water or albumen on the surface. Rolling the yolk on the filter paper many times resulted in an almost complete elimination of water and albumen. Step 3: A plastic pipette tip was used to puncture the vitelline membrane, and the yolk was collected in a 15 ml Falcon tube. Step 4: The yolk was centrifuged \cite{anton2013egg} at 7197 $g$ for $\approx$ 4 days to separate the yolk-plasma from the yolk-granules. During centrifugation, the heavier entities in the yolk (called yolk-granules) settled, leaving behind a yellow-translucent liquid known as egg yolk-plasma (see Fig.~{\ref{fig:Experiment_schematic}A}). Step 5: The total dry mass content of the yolk-plasma was determined by complete water removal (dried under fume hood) and subsequent weighing of the residue, yielding a value of 49 wt$\%$. Based on prior compositional analyses, LDLs account for approximately 85 wt$\%$ of the dry matter \cite{anton2013egg}, indicating that undiluted yolk-plasma contains $\approx$ 814 mg/ml of LDLs.

To prepare samples of lower LDL concentration, the yolk-plasma was diluted to 90 wt$\%$ and 80 wt$\%$ using NaCl buffer (170 mM), resulting in final LDL concentrations of 668 mg/ml and 547 mg/ml, respectively. The exact amounts of buffer added for each dilution are listed in Table~\ref{tab:1}.

All samples were stored at 5 $^\mathrm{o}$C until use. For XPCS measurements, the yolk-plasma solutions were loaded into 1.5 mm-diameter quartz-silica capillaries and vacuum-sealed with epoxy glue. All XPCS experiments were conducted at a temperature of 25 $^\mathrm{o}$C.

\subsection*{Estimation of $S(q)$ and isothermal compressibility $\kappa$}

The scattering intensity provided in this paper, $I(q) = I_\mathrm{raw}(q) - I_\mathrm{water}(q)$, where  $I_\mathrm{raw}(q)$ and  $I_\mathrm{water}(q)$  are the scattering intensities of sample and water, respectively, were obtained by azimuthal integration of the scattering pattern from the 2D detector. In yolk-plasma, the LDLs are dispersed in water, hence, the water background is subtracted to get the scattering contribution of LDLs. Normalization of $I(q)$ is performed such that the high-$q$ scattering intensities of higher concentrations are made to coincide with those of the $c$ = 4 mg/ml sample. The effective structure factor of the sample is estimated using, 
\begin{equation}
    S(q) = \frac{I(q)}{I_\mathrm{ff}(q)},
\label{eq:S_q}
\end{equation}
\noindent where $I_\mathrm{ff}(q)$ is the effective form factor (see Fig.~\ref{figSI:ff_fit}B-D).

\noindent The isothermal compressibility of LDL particles is estimated by,

\begin{equation}
    \kappa = \frac{S(q \to 0)} {\rho k_\mathrm{B} T}
\label{eq:I_C}
\end{equation}
where $\rho$, $k_\mathrm{B}$, and $T$ are the number density, Boltzmann constant, and sample temperature, respectively. The experimental $S(q \to 0)$ is estimated by a linear extrapolation of $S(q)$ in a $q$-range of 0.06--0.08 nm$^{-1}$ to $q=0$. The number density is estimated by dividing the volume fraction by the volume of a particle $4\pi R^{3}/3$ with $R$ being the radius obtained from the form factor fits given in Fig.~\ref {figSI:ff_fit}B-D. 

\subsection*{Estimation of $D_{0}^\mathrm{d}$ and $D_{0}$}

The diffusion coefficient of LDL particles in the dilute limit is estimated by the Stokes-Einstein relation,
\begin{equation}
    D_{0}^\mathrm{d} = \frac{k_{B} T}{6\pi \eta R},
\end{equation}

\noindent where $T$ = 298 K, $\eta$ and $R$ = 15.5 nm are the temperature of the system, viscosity of water, and radius of the LDL particles in dilute solution, respectively. However, $D_{0}$ is the effective diffusion coefficient of LDL particles in dilute solution estimated using the Stokes-Einstein relation, considering the effective radius of LDL particles at different concentrations (Fig.~\ref{figSI:ff_fit}B-D). 

\subsection*{X-ray experimental parameters}
XPCS measurements were conducted at the MID instrument \cite{madsen2021materials} of the EuXFEL in a SAXS configuration. The experiments utilized the full self-amplified spontaneous emission (SASE) with a mean photon energy of 10 keV and a focused beam size of $11.7 \pm 0.3$ µm.
The EuXFEL delivers X-ray pulse trains at a repetition rate of 10 Hz. Each train in the experiment contained 310 individual pulses with an intra-train repetition rate of 4.5 MHz, corresponding to a pulse spacing of 222 ns. This configuration allows for a maximum accessible delay time of approximately 69 µs per train, enabling access to fast dynamics in the measured correlation functions.

Scattered X-rays were recorded using the Adaptive Gain Integrating Pixel Detector (AGIPD), featuring a pixel size of  200 µm and positioned 7.68 m downstream from the sample. The primary dataset was acquired during experiment number 5397, while additional supporting data were collected during experiment 6996, which employed slightly different beam parameters. Full details of both experimental configurations are provided in the SI.

To minimize beam-induced effects, the X-ray intensity was carefully controlled using silicon single crystals and chemically vapor-deposited (CVD) diamond attenuators of varying thicknesses. Additionally, to avoid cumulative sample damage, the sample was continuously translated perpendicular to the beam at a speed of 0.4 mm/s. This ensured that each XPCS measurement was performed on a fresh volume of the sample, with negligible motion during the duration of a single pulse train.

The SAXS experimental parameters at ESRF (beamline ID02) are provided in Table.~\ref{tabSI:esrf}.

\subsection*{Hydrodynamic function}

The short-time hydrodynamic function can be decomposed into a self-part $\frac{D_\mathrm{s}^\mathrm{short}}{D_{0}}$ and a $q$-dependent part $H_\mathrm{d}(q)$ using, 

\begin{equation}
    H(q) = \frac{D_\mathrm{s}^\mathrm{short}}{D_{0}} + H_\mathrm{d}(q)
\label{eq:Hq}
\end{equation}

\noindent where $D_\mathrm{s}^\mathrm{short}$ is the self-diffusion coefficient and the $q$-dependent part of hydrodynamic function $H_\mathrm{d}(q)$ is given by \cite{nagele1996dynamics},

\begin{equation}
    H_\mathrm{d}(q) = \frac{3}{2 \pi}  \int_{0}^{\infty} \biggl(\frac{\mathrm{sin}(R\,q')}{(R\,q') } \biggr)^2  \frac{1}{1+\phi\, S_{\gamma} (R\,q')}  \,\mathrm{d}(Rq')   \times \int_{-1}^{1} \mathrm{d}x (1-x^2) [S(|q-q'|)-1]
\label{eq:Hdq}
\end{equation}

\noindent where $x$ is the cosine of the angle between the wave vectors $q$ and $q'$ and $S_{\gamma}$ is a known function independent of particle correlation and is given in ref.~\cite{genz1991collective}. For $q \to \infty$, $H_\mathrm{d}(q)$ goes to zero and $H(q)$ becomes the normalized short-time translational self-diffusion coefficient $\frac{D_\mathrm{s}^\mathrm{short}}{D_{0}}$. The $q$-dependent part $H_\mathrm{d}(q)$ was estimated using the Jscatter \cite{biehl2019jscatter} Python library, where the experimental effective structure factor of the samples was given as input.

\subsection*{Theoretical predictions of ${D_\mathrm{s}^\mathrm{short}}/{D_{0}}$ }

The theoretical prediction of $\frac{D_\mathrm{s}^\mathrm{short}}{D_{0}}$ by M. Tokuyama and I. Oppenheim \cite{tokuyama1995theory}, incorporating both short-range and long-range hydrodynamic interactions and their coupling, is given by,
\begin{equation}
\begin{gathered}
\frac{D_\mathrm{s}^\mathrm{short}}{D_{0}} = \frac{1}{1+L(\phi)}\\
L(\phi) = \frac{2b_f^{2}}{1-b_f} - \frac{c_f}{1+2c_f} + \biggl[ -\frac{2b_f c_f}{1-b_f+c_f} 
\biggl( 1-\frac{6 b_f c_f}{1-b_f+c_f+4b_f c_f} + \frac{2b_f c_f}{1-b_f+c_f+2b_f c_f} \biggr)\\
+ \frac{b_f c_f^{2}}{(1+c_f)(1-b_f+c_f)}\biggl( 1+\frac{3b_fc_f^{2}}{(1+c_f)(1-b_f+c_f)-2b_fc_f^{2}} - \frac{b_f c_f^{2}}{(1+c_f)(1-b_f+c_f)-b_f c_f^{2}}\biggr) \biggr]\\
b_f = (9\phi/8)^{1/2}, \; c_f = 11\phi/6
\end{gathered}
\label{eq:Ds_D0_tokuyama}
\end{equation}

\noindent The first, second, and third terms in $L(\phi)$ correspond to contributions from the long-range hydrodynamic interaction, short-range hydrodynamic interaction, and their coupling, respectively \cite{tokuyama1995theory}.

\subsection*{Modeling of MSD}
An expression to mean-squared displacement for concentrated, equilibrium suspensions of hard-spheres is given by Tokuyama et al. \cite{tokuyama2001slow,tokuyama2003test},

\begin{equation}
  \mathrm{MSD}\, (t) = \frac{1}{\nu}  \; \mathrm{ln} \Bigg[1+ \frac{D_\mathrm{s}^\mathrm{short}}{D_\mathrm{s}^\mathrm{long}} \;\;\mathrm{exp} \,(6 \nu D_\mathrm{s}^\mathrm{long} t)\Bigg]
\label{eq:msd_fit}
\end{equation}
\noindent where $\nu$ is a free parameter to be determined and is related to the static properties in the equilibrium suspension \cite{tokuyama2001slow}.

\subsection*{Assessment of X-ray beam-induced effects}

In XPCS measurements, X-ray beam-induced damage is characterized by the changes in the native structure and equilibrium dynamics \cite{marioNatComm}. To determine the critical dose and dose rate for yolk-plasma, we performed XPCS measurements at various incident X-ray intensities such that the dose rate on the sample is different for the same exposure time. The results of beam-induced changes in the structure and dynamical parameters of yolk-plasma are given in the SI. The yolk-plasma was found to be stable until a critical dose of 20 kGy, and hence all measurements shown in this manuscript are performed below the critical threshold of 20 kGy using a dose rate of 0.23 kGy/µs. During the measurement, X-ray-induced heating led to an approximate temperature increase of 2 K in the sample (see SI for further details).

\begin{figure} 
	\centering
	\includegraphics[width=0.95 \textwidth]{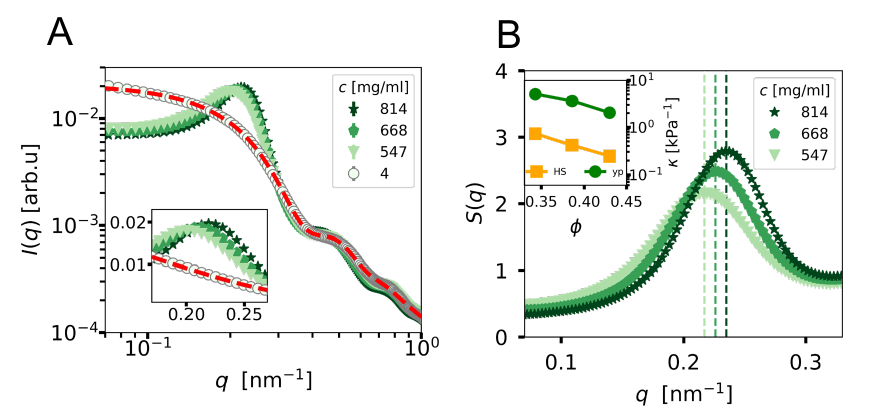} 
	\caption{\textbf{Small angle X-ray scattering of yolk-plasma at different concentrations: } (\textbf{A}) Normalized scattering intensity of egg yolk-plasma at different concentrations, obtained at ESRF. The red dashed line is the spherical core-shell fit to dilute $I(q)$ data ($c$ = 4 mg/ml) with radius $R = $ 15.5 nm, including generalised Guinier-Porod contribution (see SI for more details).
    A zoomed-in view around the structure factor peak $q$-regime is provided in the inset.
    (\textbf{B}) The effective structure factor, $S(q)$, is extracted by dividing $I(q)$ by the respective form factors given in Fig.~\ref{figSI:ff_fit}B-D. Inset: isothermal compressibility (Eq.~\ref{eq:I_C}) as a function of volume fraction for yolk-plasma samples compared with hard-spheres. 
}
	\label{fig:Iq_Sq} 
\end{figure}

\begin{figure} 
	\centering
	\includegraphics[width=0.95 \textwidth]{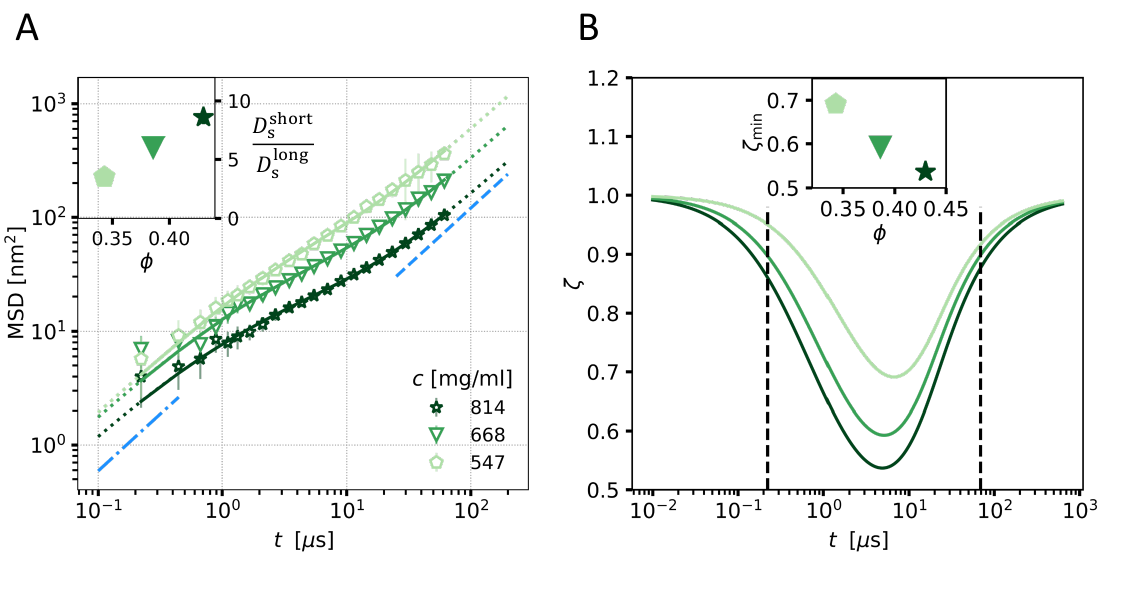} 
	\caption{\textbf{Mean squared displacement and the anomalous exponent:}  (\textbf{A}) Comparison of MSD of all samples. The solid curves are fits using Eq. \ref{eq:msd}. The dotted curves are the extrapolation of the fits. The blue dash-dotted and dashed lines indicate MSD $\sim t$. These lines serve as a guide to the eye and indicate the short-time and long-time behavior of the MSD, respectively. The inset shows the ratio $D_\mathrm{s}^\mathrm{short}/D_\mathrm{s}^\mathrm{long}$ obtained from the fit as a function of volume fraction.
    (\textbf{B}) Anomalous exponent extracted using Eq. \ref{eq:zeta}, as a function of time. The dashed vertical line indicates the experimentally investigated time scale of 0.22--69 µs. The inset shows the minimum value of $\zeta$ as a function of volume fraction.
}
	\label{fig:MSD} 
\end{figure}


\clearpage 



\section*{Acknowledgments}

We acknowledge European XFEL in Schenefeld, Germany, for the provision of X-ray free-electron laser beamtime at the Scientific Instrument MID (Materials Imaging and Dynamics) and would like to thank the staff for their assistance. The data presented here were taken as part of the beamtimes with experiment numbers 5397 and 6996. We acknowledge the participation of Marvin Kowalski, Mohammad Sayed Akhundzadeh, Jenny Schrage, Pia Kwiasowski, Maximilian Senft, Lara Reichart, and Chang Hee Woo during the EuXFEL beamtime. We thank the European Synchrotron Radiation Facility (ESRF) for the provision of synchrotron radiation facilities at TRUSAXS instrument (Time Resolved Ultra Small Angle X-ray Scattering), ID02 (SC-5435), with additional support from Theyencheri Narayanan. We acknowledge the participation of Randeer Gautam, Moritz Puritscher, Fabian Prandl, Lovisa Jansson, Milla Åhlfeldt, and Sampad Bag during the ESRF beamtime. We acknowledge financial support by the consortium DAPHNE4NFDI in association with the German National Research Data Infrastructure (NFDI) e.V. - project number 4602487. This work was supported through the Maxwell computational resources operated at Deutsches Elektronen-Synchrotron DESY, Hamburg, Germany. We acknowledge Dr. D.C.F. Wieland for providing access to the Bolin Gemini rotational HR nano-rheometer for rheology measurements. M.P. thanks the DELTA machine group for providing synchrotron radiation for sample characterization. 

\paragraph*{Funding:}

We acknowledge BMBF (05K19PS1, 05K20PSA and 05K22PS1)(to C.G.); 05K19VTB, (to F.S. and F.Z.), DFG-ANR (SCHR700/28-1, SCHR700/42-1, ANR-16-CE92-0009) (to F.S. and F.Z.). C.G. acknowledges funding from NFDI 40/1 (DAPHNE4NFDI). F.P., I.A.A., and A.G. acknowledge funding from the European Union’s Horizon Europe research and innovation programme under the Marie Skłodowska-Curie grant agreement No. 101081419 (PRISMAS) (F.P. and I.A.A.) and 101149230 (CRYSTAL-X) (F.P. and A.G.). F.P. acknowledges financial support by the Swedish National Research Council (Vetenskapsrådet) under Grant No. 2019-05542, 2023-05339 and within the Röntgen-Ångström Cluster Grant No. 2019-06075, and the kind financial support from Knut och Alice Wallenberg foundation (WAF, Grant. No. 2023.0052). This research is supported by the Center of Molecular Water Science (CMWS) of DESY in an Early Science Project, the MaxWater initiative of the Max-Planck-Gesellschaft (Project No. CTS21:1589), Carl Tryggers and the Wenner-Gren Foundations (Project No. UPD2021-0144).

\paragraph*{Author contributions:}
N.D.A, M.D., and C.G. designed the experiment, along with discussions with F.S., F.P., F.Z, M.P., and F.L.. N.D.A. and M.D. prepared the yolk-plasma samples. N.D.A. planned the measurements at EuXFEL. N.D.A., M.D., S.T., A.G., S.R., A.M.R., A.L., F.U., M.B., P.P.R., I.A., F.L., M.P., and C.G. conducted the experiment. J.M., W.J., J.H., A.R., J.P., F.B., U.B., M.Y., R.S., R.R., and A.M. operated the MID instrument at EuXFEL. J.W., A.L., F.B., and J.M. provided the MID data analysis pipeline. N.D.A. performed the data processing and analysis. N.D.A., M.D., C.G., F.Z., and F.S. discussed the XPCS data analysis with input from A.G., F.P., J.M., F.L., and M.P.. W.C. operated the ID02 beamline at ESRF. The manuscript was written by N.D.A., and C.G. with input from all authors.

\paragraph*{Competing interests:}
There are no competing interests to declare.

\paragraph*{Data and materials availability:}
The processed data (scattering intensity profiles, intermediate scattering functions) of yolk-plasma samples have been deposited in Zenodo (\url{xxxx}).
Any other data used in this study are available from the authors upon request. Custom Python scripts developed for the study are available at Zenodo (\url{xxxx}).

Data recorded for the experiment at the European XFEL are available at [doi:10.22003/XFEL.EU-DATA-005397-00].
Data recorded at ESRF are available at [doi:10.15151/ESRF-ES-2009918726].

\subsection*{Supplementary materials}
Supplementary Text\\
Figs. S1 to S16\\
Tables S1 to S3\\


\newpage


\renewcommand{\thefigure}{S\arabic{figure}}
\renewcommand{\thetable}{S\arabic{table}}
\renewcommand{\theequation}{S\arabic{equation}}
\renewcommand{\thepage}{S\arabic{page}}
\setcounter{figure}{0}
\setcounter{table}{0}
\setcounter{equation}{0}
\setcounter{page}{1} 


\begin{center}
\section*{Supplementary Materials for\\ \scititle}

\author{
    Nimmi Das Anthuparambil$^{\ast}$,
    Michelle Dargasz,
    Sonja Timmermann, \and
    Anita Girelli,
    Sebastian Retzbach,
    Johannes Möller,
    Wonhyuk Jo, \and
    Agha Mohammad Raza,
    Aliaksandr Leonau,
    James Wrigley, \and
    Frederik Unger,
    Maddalena Bin,
    Prince Prabhu Rajaiah,
    Iason Andronis, \and
    William Chèvremont,
    Jörg Hallmann,
    Angel Rodriguez-Fernandez,
    Jan-Etienne Pudell, \and
    Felix Brausse,
    Ulrike Boesenberg, 
    Mohamed Youssef,
    Roman Shayduk, \and
    Rustam Rysov,
    Anders Madsen,
    Felix Lehmkühler,
    Michael Paulus, \and
    Fajun Zhang,
    Fivos Perakis,
    Frank Schreiber,
    Christian Gutt $^{\dagger}$ \and
\small$^\ast$Corresponding author. Email: nimmi.das.anthuparambil@desy.de\and
\small$^\dagger$Corresponding author. Email: christian.gutt@uni-siegen.de\and}

\end{center}

\subsubsection*{This PDF file includes:}

Supplementary Text\\
Figures S1 to S16\\
Tables S1 to S3\\

\newpage

\section{Sample details}\label{yolk_details}

\begin{figure} 
	\centering
	\includegraphics[width=0.95 \textwidth]{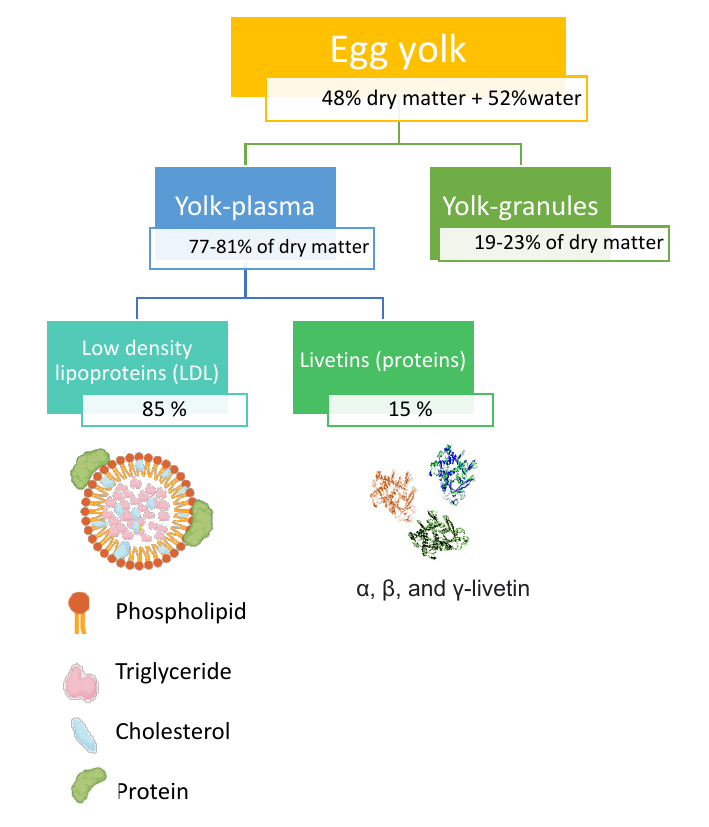} 
	\caption{\textbf{Constituents of hen egg yolk:} The egg yolk can be separated into two major fractions: egg yolk-plasma and egg yolk-granules via moderate centrifugation as described in the text. }
\label{yolk_contents}
\end{figure}

Egg yolk is a complex assembly of proteins, phospholipids, and cholesterol in water. The egg yolk constituents are described in Fig. \ref{yolk_contents}. Yolk consists of approximately 50$\%$ water, 33$\%$ lipid, and 17$\%$ protein\cite{mann2008chicken}. This complex natural assembly is stabilized via interactions between its constituents. Egg yolk can be separated into two fractions (plasma and granules) by moderate centrifugation \cite{anton2013egg}. The egg yolk-plasma is a translucent liquid which represents $\approx$ $ 80\%$ of yolk dry matter. The major constituent of egg yolk-plasma is low-density lipoproteins (LDLs) ($85\%$ of yolk-plasma), which are spherical core-shell nanoparticles \cite{anton2013egg}. The core is made of triglycerides and cholesterol esters in a liquid state, and the shell is made of a monolayer of phospholipids onto which apolipoproteins are attached, as shown in Fig. \ref{yolk_contents}. The LDLs are water-soluble entities, and the density of LDL is 0.982 g/cm$^{3}$. The hydrophobic interactions between the phospholipid tail and core lipids and hydrophilic interactions between the phospholipid head and water molecules stabilize this complex nano-assembly. LDLs contain 11-17$\%$ proteins and 83-89$\%$ lipids (neutral lipids and phospholipids) \cite{anton2013egg}.

\clearpage 
\section{SAXS data}
\begin{figure} 
	\centering
	\includegraphics[width=0.6 \textwidth]{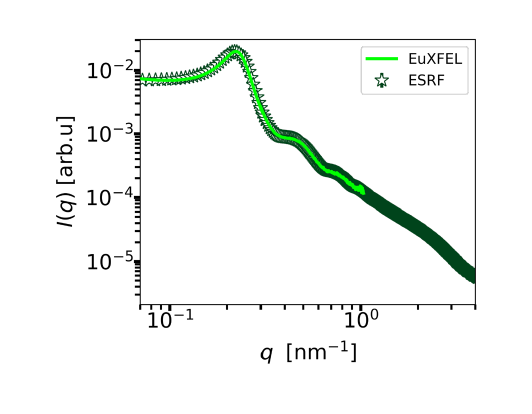} 
	\caption{\textbf{SAXS from ESRF and EuXFEL:} The $I(q)$ of pure yolk-plasma  ($c$ = 814 mg/ml) obtained from EuXFEL and ESRF. }
\label{figSI:saxs_xfel_esrf}
\end{figure}

The SAXS data collected from EuXFEL and ESRF are in good agreement, as shown in Fig.~\ref{figSI:saxs_xfel_esrf}.

\subsection{Form factor modeling of yolk-plasma}

\begin{figure} 
	\centering
	\includegraphics[width=0.9 \textwidth]{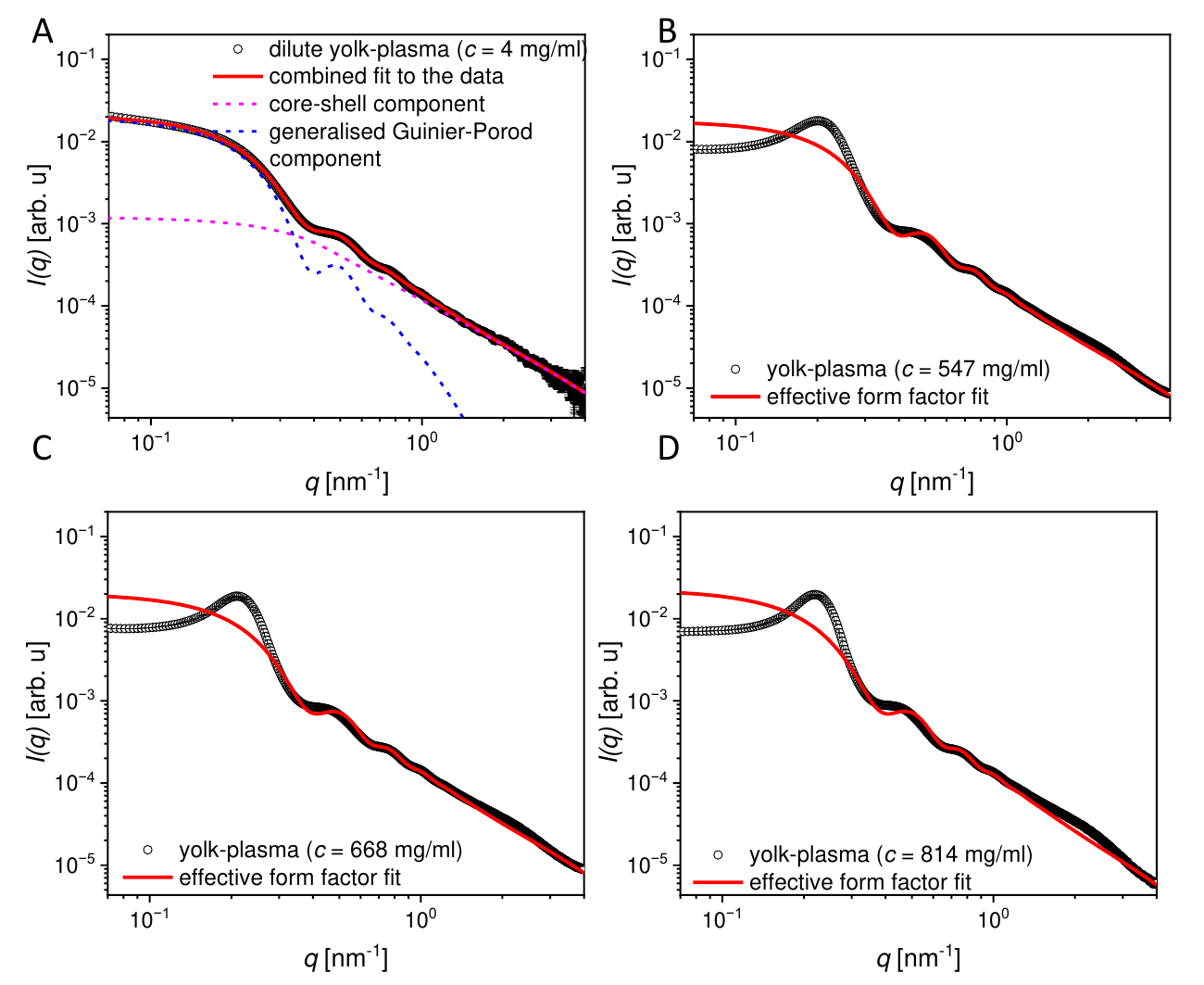} 
	\caption{\textbf{Form factor of yolk-plasma:} \textbf{(A)} The scattering intensity of yolk-plasma ($c = 4$ mg/ml) along with a theoretical model fit (red). The fit contains contributions of the core-shell form factor and a generalised Guinier-Porod contribution. The $I(q)$ of \textbf{(B)} $c = 547$ mg/ml \textbf{(C)} $c = 668$ mg/ml and \textbf{(D)} $c = 814$ mg/ml along with a theoretical model fit.    
}
\label{figSI:ff_fit} 
\end{figure}

\begin{figure} 
	\centering
	\includegraphics[width=0.8 \textwidth]{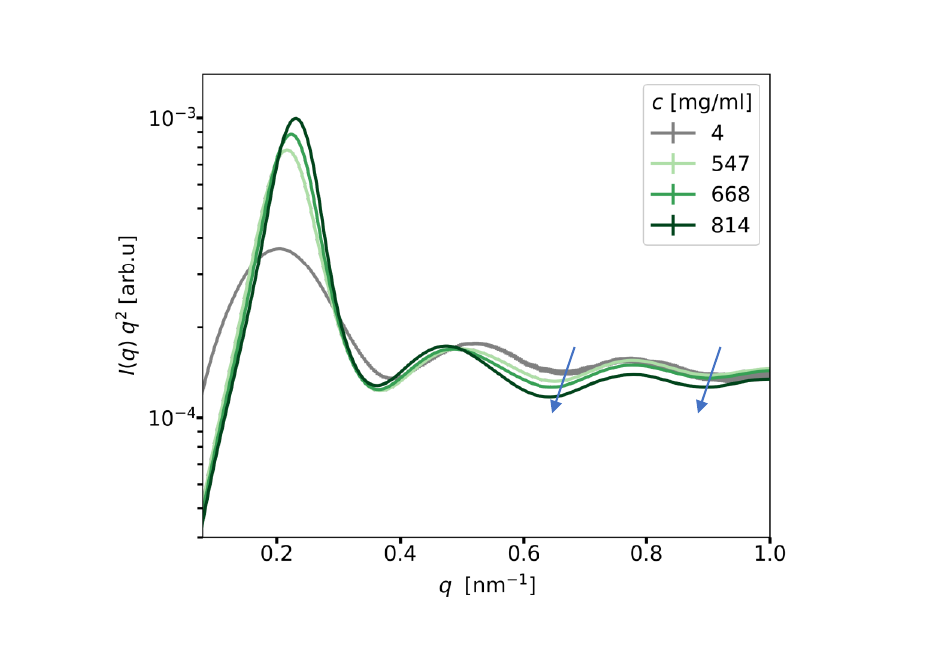} 
	\caption{\textbf{Kratky plot of yolk-plasma:} The kratky plot ($q^2I(q)$ vs $q$) of all samples. The blue arrows indicate the shift in minima positions with increasing concentration.
}
\label{figSI:kratky} 
\end{figure}

For modeling the form factor of yolk-plasma, the SASfit software \cite{bressler2015sasfit} was used. The fit is comprised of a core-shell component plus a generalized Guinier-Porod component. The core-shell component is realized by using "ExpShell" model in SASfit software, where the shell has an exponentially varying contrast profile. A generalized Guinier-Porod component is implemented to account for the presence of apolipoprotein on the LDL corona \cite{cao2021protein}. The SAXS data obtained from ESRF is used for formfactor modeling due to the extended $q$ range and lower noise levels at higher $q$ values (Fig.~\ref{figSI:saxs_xfel_esrf}). First, experimental data of $c =$ 4 mg/ml was fitted using the combined model as depicted in Fig.~\ref{figSI:ff_fit}A. The Schulz-Zimm distribution of radius with a core radius of $\approx$ 12 nm and a shell thickness of $\approx$ 3.5 nm is used for the core-shell model. The dispersity (standard deviation/mean radius) was estimated as 0.126. As the concentration increases, a change in the size of LDL particles is expected due to their soft, deformable nature and the limited space available, consistent with prior observations of lipid vesicles \cite{quinn2022crowding} and micelles \cite{ogino1988micelle}. As depicted in Fig.~\ref{figSI:kratky}, a small shift in the second and third minima with increasing concentration indicates the change in the size of LDLs. Hence, we estimate the effective form factor at high concentrations by modeling $I(q)$ at $q > 0.65$ nm$^{-1}$. For this, the model fit given in Fig.~\ref{figSI:ff_fit}A is used as an initial guess for other samples (Fig.~\ref{figSI:ff_fit}B-D). The resulting form factor fits for all sample cases are given in Fig.~\ref{figSI:ff_fit}. The estimated decrease in particle size in the highest concentrated sample ($c = 814$ mg/ml) is $\approx 3\%$ compared to the dilute case ($c = 4$ mg/ml). The effective radius of LDL particles determined from form factor modeling is subsequently utilized for the estimation of other parameters, such as $D_{0}$ (see Methods and Materials section), $\kappa$ of hard-spheres (given in the inset of Fig.~\ref{fig:Iq_Sq}B), $k_BT/(6\pi R D_\mathrm{s}^\mathrm{long})$ given in Fig.~\ref{figSI:viscosity} and storage modulus estimate of hard-spheres given in Fig.~\ref{figSI:moduli}C. Note that experimental estimation of the hydrodynamic radius at high concentrations was difficult using a dynamic light scattering technique due to multiple scattering issues, and hence, for the estimation of $D_{0}$, the radius obtained from SAXS is used.

\subsection{Effective $S(q)$ }

\begin{figure} 
	\centering
	\includegraphics[width=0.8 \textwidth]{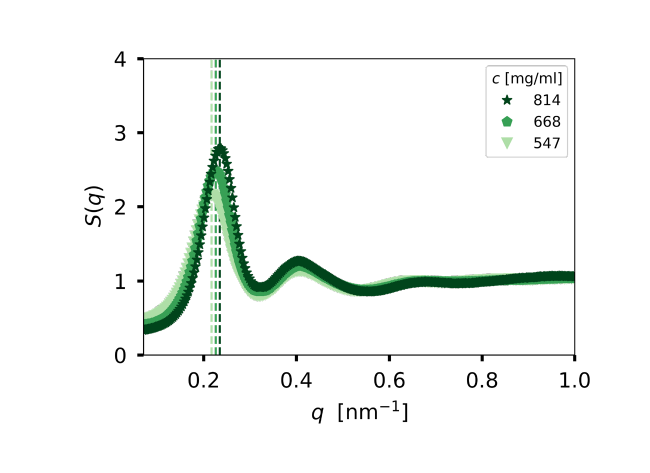} 
	\caption{\textbf{Effective $S(q)$:} The structure factor estimated using the respective form factors given in Fig.~\ref{figSI:ff_fit}B-D.
}
\label{figSI:full_Sq} 
\end{figure}

The effective structure factor of yolk-plasma at different concentrations is estimated using Eq.~\ref{eq:S_q}, experimental $I(q)$, and form factor fits given in Fig.~\ref{figSI:ff_fit}B-D. The extracted $S(q)$ for all samples is given in Fig.~\ref{figSI:full_Sq}.

\section{Modeling of correlation functions from XPCS}

\begin{figure} 
	\centering
	\includegraphics[width=0.6 \textwidth]{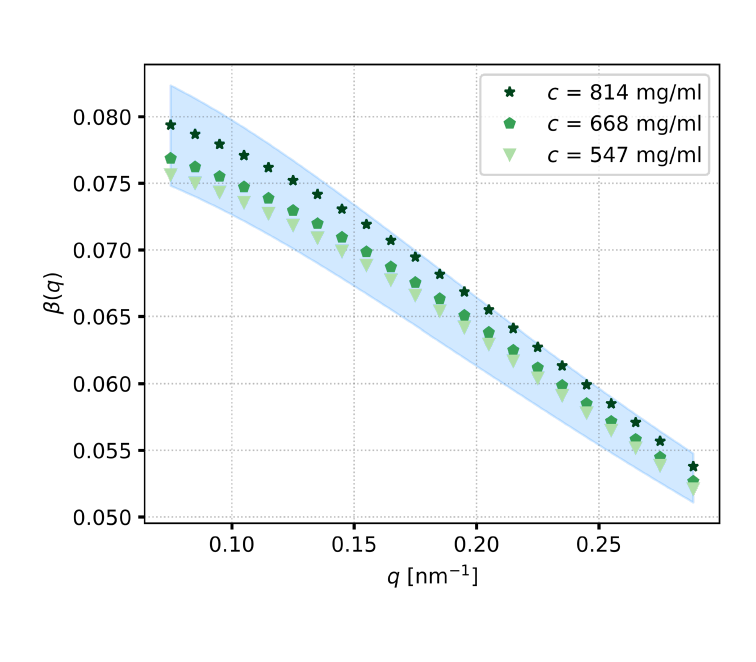} 
	\caption{\textbf{Speckle contrast:} The blue region indicates the speckle contrast estimated using the relationship given in Eq.~\ref{eqSI:contrast} and ~\ref{eqSI:contrast_M} for a beamsize of $b = 11.7\pm0.3 \,$µm. The data points indicate the speckle contrast obtained from the fit of yolk-plasma samples.}
\label{figSI:speckle_contrast} 
\end{figure}

The intensity autocorrelation functions provided in the main manuscript extracted from XPCS measurements are modeled using Eq.~\ref{eq:g2}. The speckle contrast was estimated using the relationship provided in ref.~\cite{hruszkewycz2012high},

\begin{equation}
    \beta(q) = \beta_\mathrm{t}(q), \beta_\mathrm{l}(q)
\label{eqSI:contrast}
\end{equation}
\noindent where $ \beta_\mathrm{t}(q)$ is the effective speckle contrast due to the transverse
coherence. $ \beta_\mathrm{t}(q) \approx 0.5$ was found for European XFEL \cite{marioNatComm}. $\beta_\mathrm{l}(q)$ describes the loss of coherence due to energy bandwidth, experiment geometry, and speckle shape and is given by $\beta_\mathrm{l}(q) = 1/(M_\mathrm{rad} M_\mathrm{det})$ with, 

\begin{equation}
\begin{gathered}
    M_\mathrm{rad} = \sqrt{1+  \frac{q^2 (dE/E)^2 \;(b^2 \mathrm{cos}^2(\upvartheta) + d^2 \mathrm{sin}^2(\upvartheta)}{4 \pi^2}}\\
    M_\mathrm{det} = \sqrt{1+  \frac{p^4 b^2 (b^2 \mathrm{cos}^2(2 \upvartheta) + d^2 \mathrm{sin}^2(2 \upvartheta)}{\lambda^4 L^4 M_{rad}^2}}
\end{gathered}
\label{eqSI:contrast_M}
\end{equation}

\noindent where $ dE/E \approx 10^{- 2}$, $b = 11.7\pm0.3$ µm, $d = 1.5$ mm, $p$ = 200 µm, $\lambda$, $L$ and $2 \upvartheta$ are the effective energy bandwidth, beam size, thickness of capillary, pixel size, wavelength of X-rays, sample to detector distance, and scattering angle, respectively. The effective energy bandwidth is estimated by modeling $\beta(q)$ of ludox power samples using Eq.~\ref{eq:g2}. The values of other parameters can be found in Table~\ref{tabSI:beamtime}. The blue shaded region in Fig.~\ref{figSI:speckle_contrast} indicates the confidence interval for contrast values (beamtime id: p005397). The fitted contrast of all samples lies within this confidence interval.

\clearpage

\section{Estimation of $\Gamma^{s} (q)$}

\begin{figure} 
	\centering
	\includegraphics[width=0.5 \textwidth]{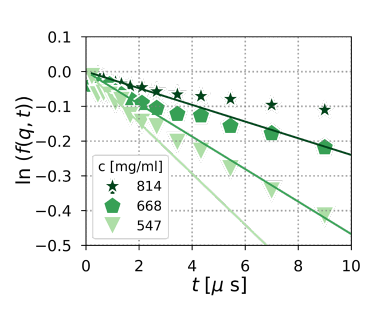} 
	\caption{The exponential fit (solid line) of intermediate scattering function $f(q,t)$ of all three concentrations at $q = 0.225\, \mathrm{nm}^{-1}$. 
}
\label{figSI:short_time_g2fit} 
\end{figure}

To estimate $\Gamma^{s} (q)$, the logarithm of intermediate scattering function $f(q,t) = ( \frac{g_2(q,t)-1}{\beta}  )^{1/2}$ was fitted using a linear function as depicted in Fig.~\ref{figSI:short_time_g2fit} for $t<$ 2 µs.

\section{$D(q)$ and $S(q)$}

\begin{figure} 
	\centering
	\includegraphics[width=0.75 \textwidth]{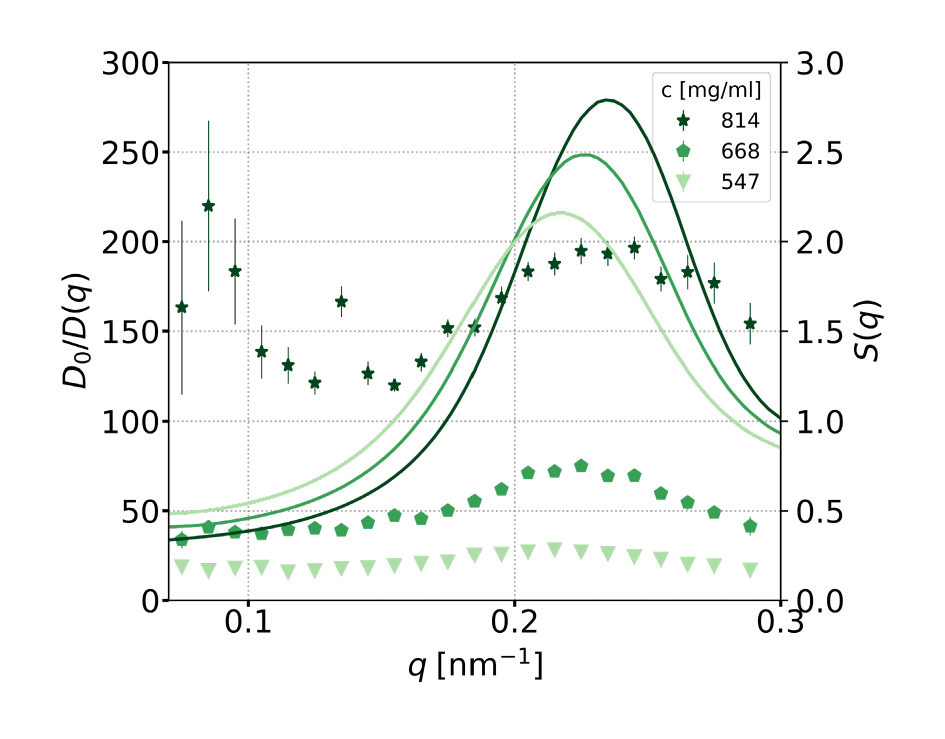} 
	\caption{\textbf{$D(q)$ and $S(q)$:} The $D_{0}/D(q)$ (solid symbols) and $S(q)$ (solid lines) are plotted in a single plot with a double Y-axis.}
\label{figSI:Sq_D0_Dq}
\end{figure}

The $D_{0}/D(q)$ and $S(q)$ of egg yolk-plasma at different concentrations are compared in a double-Y plot as shown in Fig.~\ref{figSI:Sq_D0_Dq}.

\section{Modeling of hydrodynamic interactions}

\begin{figure} 
	\centering
	\includegraphics[width=0.5 \textwidth]{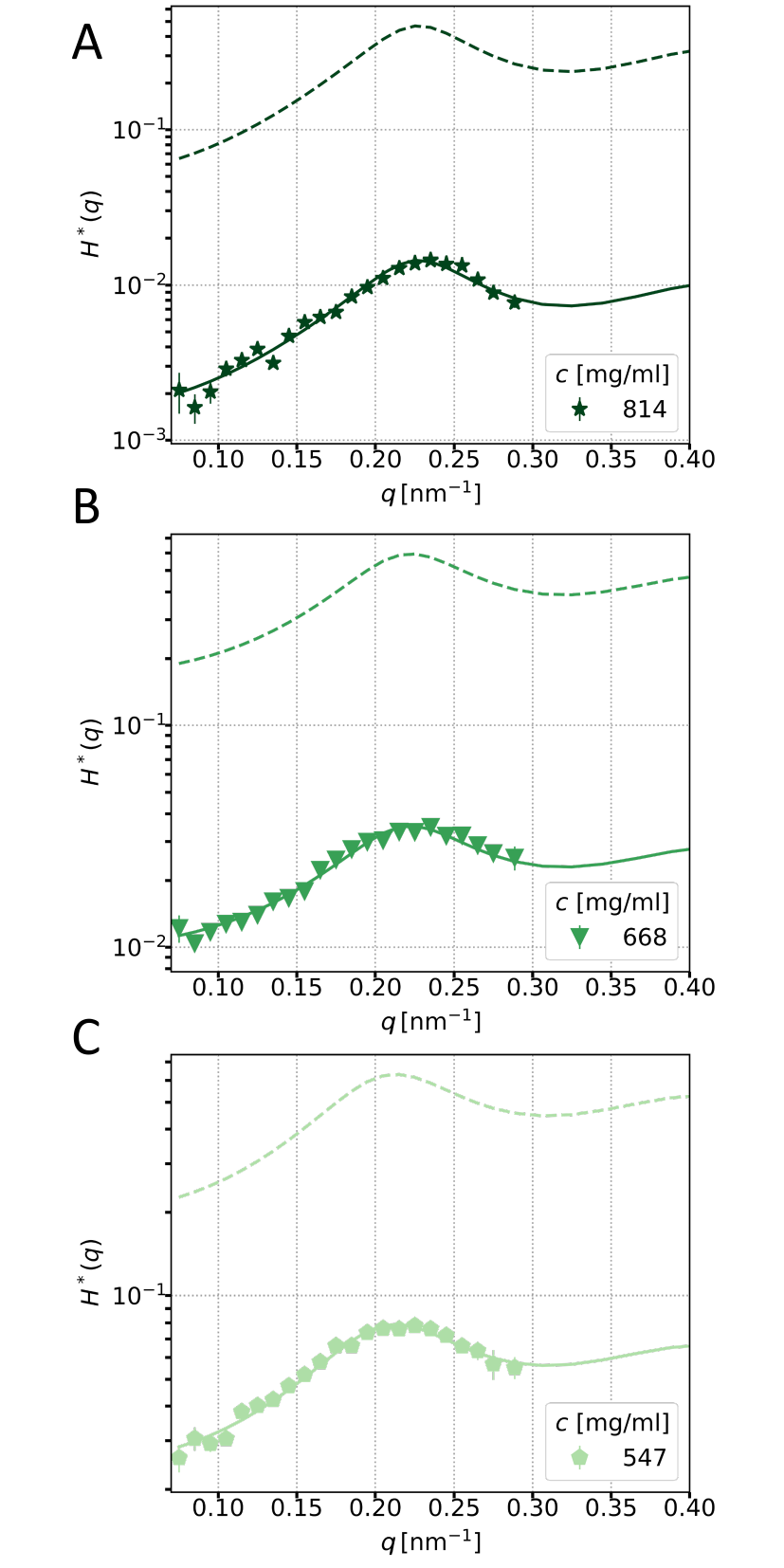} 
	\caption{\textbf{Modeling of hydrodynamic interactions:} The extracted hydrodynamic function $H^*(q)$ (data points) from experimental data for  (\textbf{A}) $c =$ 814 mg/ml, (\textbf{B}) $c =$ 668 mg/ml, and (\textbf{C}) $c =$ 547 mg/ml along with fits using Eq.~\ref{eq:Hq} (dashed lines) and rescaled hydrodynamic function (solid lines) $H^*(q)$ given in the text.
}
\label{figSI:SI_Hq_fits} 
\end{figure}

The $q$-dependence of experimentally extracted hydrodynamic functions was captured by the theoretical expressions given in the main text (Eq.~\ref{eq:Hq} and Eq.~\ref{eq:Hdq}). The $q$-dependent part $H_\mathrm{d}(q)$ was estimated using the Jscatter \cite{biehl2019jscatter} Python library, where the experimental structure factor of the samples was used as input. The rescaled hydrodynamic function fits the experimental data well (solid line in Fig.~\ref{figSI:SI_Hq_fits}). The dashed lines are theoretical predictions using Eq.~\ref{eq:Hq}. 

\clearpage
\section{Comparison of experimental $D_\mathrm{s}/D_\mathrm{0}$ with theoretical model of Medina-Noyola et al.}

\begin{figure} 
	\centering
	\includegraphics[width=0.6 \textwidth]{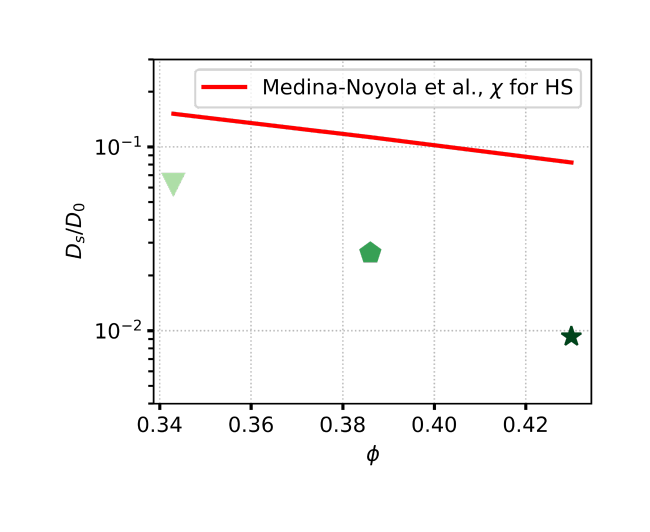} 
	\caption{\textbf{Comparison of experimental $D_\mathrm{s}/D_\mathrm{0}$ with theoretical model of Medina-Noyola et al.:} The $D_\mathrm{s}/D_\mathrm{0}$ (solid symbols) and model fit (orange) using Eq.~\ref{eqSI:medina}.}
\label{figSI:medina}
\end{figure}

The theoretical estimation of the normalized long-time self-diffusion coefficient proposed by Medina-Noyola \cite{medina1988long} and Blaaderen et al. \cite{van1992long} is,
\begin{equation}
\begin{gathered}
D_\mathrm{s}^\mathrm{long}/D_\mathrm{0} = (1 - \phi)/(1 + (3/2)\phi) / (1+2\chi\phi) \\
\chi = (1+0.5\phi)/(1-\phi)^2 \qquad \mathrm{[for \; hard-spheres]}
\end{gathered}
\label{eqSI:medina}
\end{equation}

\noindent The $D_\mathrm{s}/D_\mathrm{0}$ of egg yolk-plasma at different concentrations is compared to the model given in Eq.~\ref{eqSI:medina} as shown in Fig.~\ref{figSI:medina}.

\section{Memory Equation}

The memory equation relating $f(q,t)$  to the collective memory function $M_\mathrm{c} (q,t)$ is given by \cite{NAGELE1997PhyA},
\begin{equation}
    \frac{\partial f(q,t)}{\partial t} = - q^2 D^\mathrm{short}(q)\; f(q,t) + q^2 \int_{0}^{t} \ M_\mathrm{c} (q,t-u) \frac{f(q,u)}{S(q)} \,\mathrm{d}u 
\label{eq:memory_eq}
\end{equation}
\noindent where $u$ is the integration time parameter. The first term on the right-hand side determines the short-time dynamics, describing an exponential decay of $f(q,t)$ due to free or weakly interacting particle motion. The second term, often referred to as the memory term, relates the present rate of change of $f(q,t)$ to its past values, capturing the effects of many-body interactions and dynamic caging.  In the short-time ($t<<\tau_\mathrm{i}$) regime, the second term in Eq. \ref{eq:memory_eq} does not contribute and hence $f(q,t)$ decays exponentially. However, for $t\gtrapprox\tau_\mathrm{i}$, memory effects become significant, leading to a slower, non-exponential decay of $f(q,t)$, characteristic of particle caging and the onset of viscoelastic behavior.

\section{MSD estimation}

\begin{figure} 
	\centering
	\includegraphics[width=1 \textwidth]{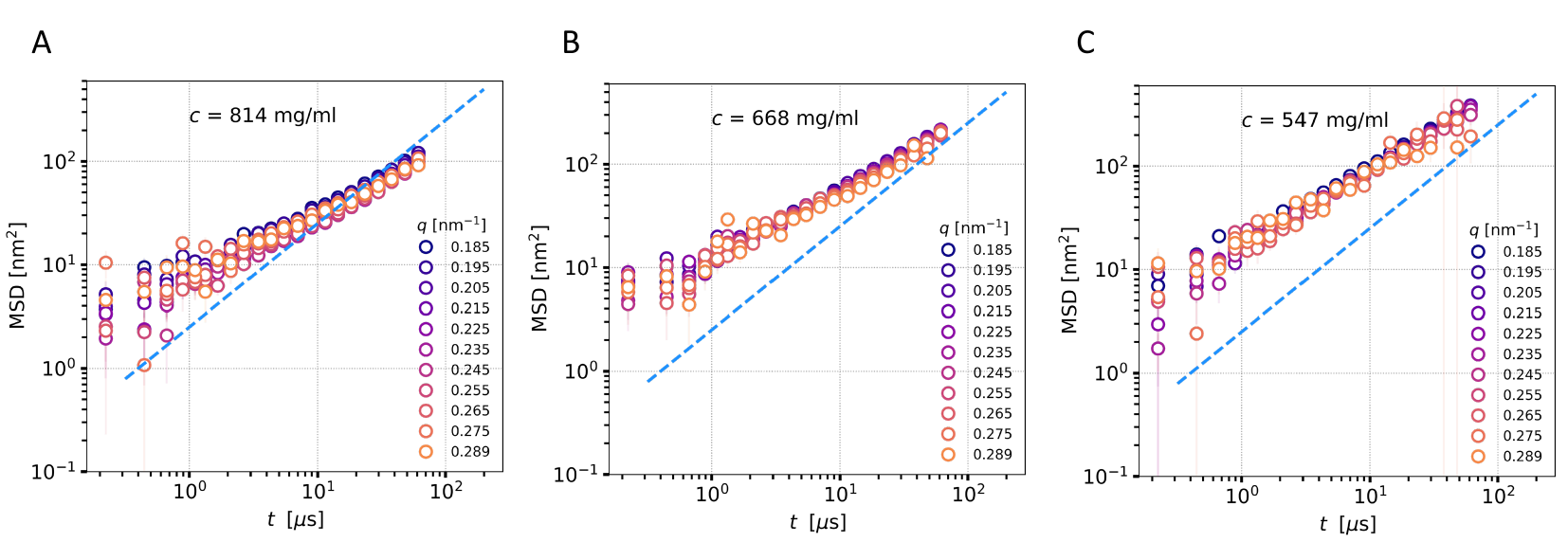} 
	\caption{\textbf{MSD at different $q$ values: } The extracted MSD of (\textbf{A}) $c =$ 814 mg/ml, (\textbf{B}) $c =$ 668 mg/ml, and (\textbf{C}) $c =$ 547 mg/ml for $q$ values in the range 0.185--0.288 nm$^{-1}$ ($q R> 2.5$). The color variation from purple to orange indicates increasing $q$ values. The dashed line indicates linear MSD $\sim t$ behavior.}
\label{figSI:MSD_diffq} 
\end{figure}

The MSD of all samples at $qR> 2.5$ is estimated using the relation Eq. \ref{eq:msd}. The calculated MSD values collapse to a single curve (Fig. \ref{figSI:MSD_diffq}), indicating the validity of Eq. \ref{eq:msd}.

\section{Viscoelastic properties from MSD}

\begin{figure} 
	\centering
	\includegraphics[width=1 \textwidth]{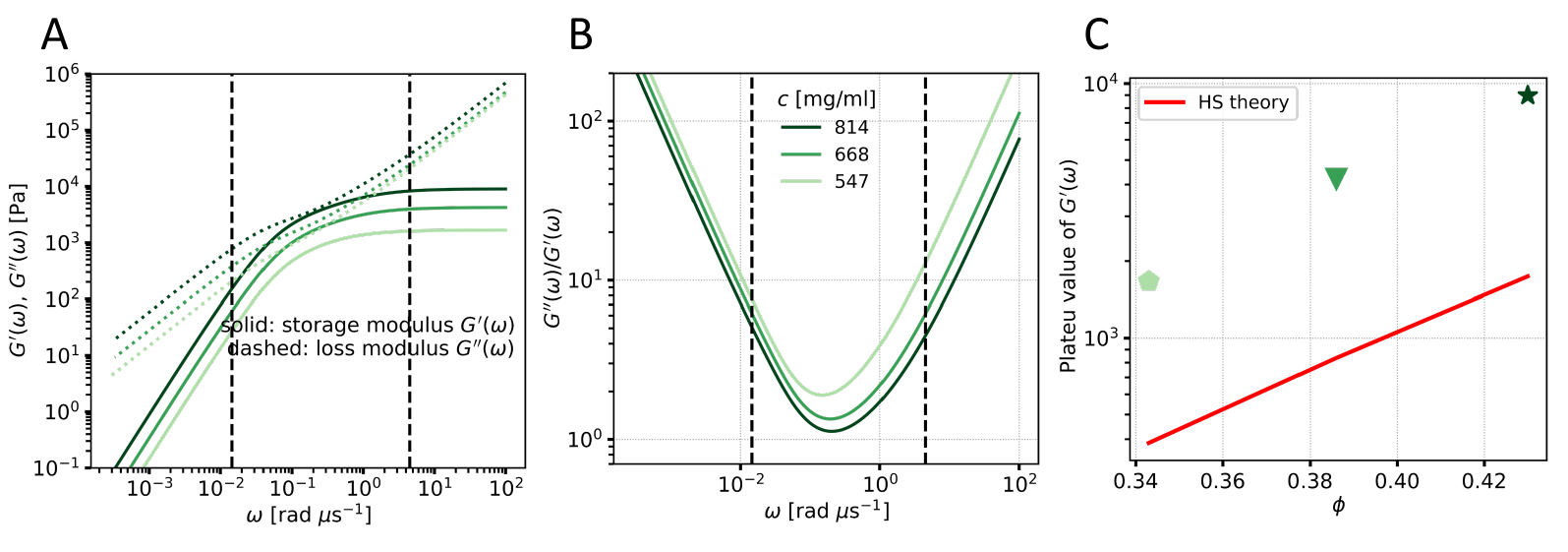} 
	\caption{\textbf{Storage and loss modulus: } (\textbf{A}) Comparison of storage (solid curve) and loss modulus (dashed curve) of all samples. (\textbf{B}) The ratio between loss modulus to storage modulus gives loss-tangent. Note that a loss-tangent greater than unity indicates that the sample is in a liquid state. The vertical dashed lines in \textbf{A} and \textbf{B} indicate the experimental time window. (\textbf{C}) Plateau value of $G'(\omega)$ as a function of volume fraction. The red curve indicates the theoretical prediction for hard-spheres (Eq.~\ref{eqSI:storage_HS}).
    }
\label{figSI:moduli} 
\end{figure}

The frequency-dependent ($\omega$) visco-elastic properties of complex fluids can be extracted from the time-dependent MSD using the formalism described in \cite{mason2000estimating}. Here, we estimate the storage modulus, $G'(\omega)$, and loss modulus, $G''(\omega)$, of all samples from the MSD fit given in Fig. \ref{fig:MSD} using the formalism provided in ref. \cite{mason2000estimating}. The estimated values are shown in Fig. \ref{figSI:moduli}A. At all times $G''(\omega) > G'(\omega)$, indicated by loss-tangent (= $G''(\omega) / G'(\omega)$) $>$1 in Fig.~\ref{figSI:moduli}B. This confirms that all samples are in the liquid state. For hard-spheres the localization length $r_\mathrm{L}$ determines the storage modulus via,
\begin{equation}
G'(\omega) = \frac{9}{5 \pi} \; \frac{\phi k_\mathrm{B}T}{2R r_\mathrm{L}^2}.
\label{eqSI:storage_HS}
\end{equation}

The estimated values of $G'(\omega)$ with $r_\mathrm{L} = r_\mathrm{cage}$ are shown in Fig.~\ref{figSI:moduli}C. Clearly, the hard-sphere model underestimates the $G'(\omega)$ values of LDL solutions.

\clearpage

\section{Viscosity from rheology}

\begin{figure} 
	\centering
	\includegraphics[width=0.9 \textwidth]{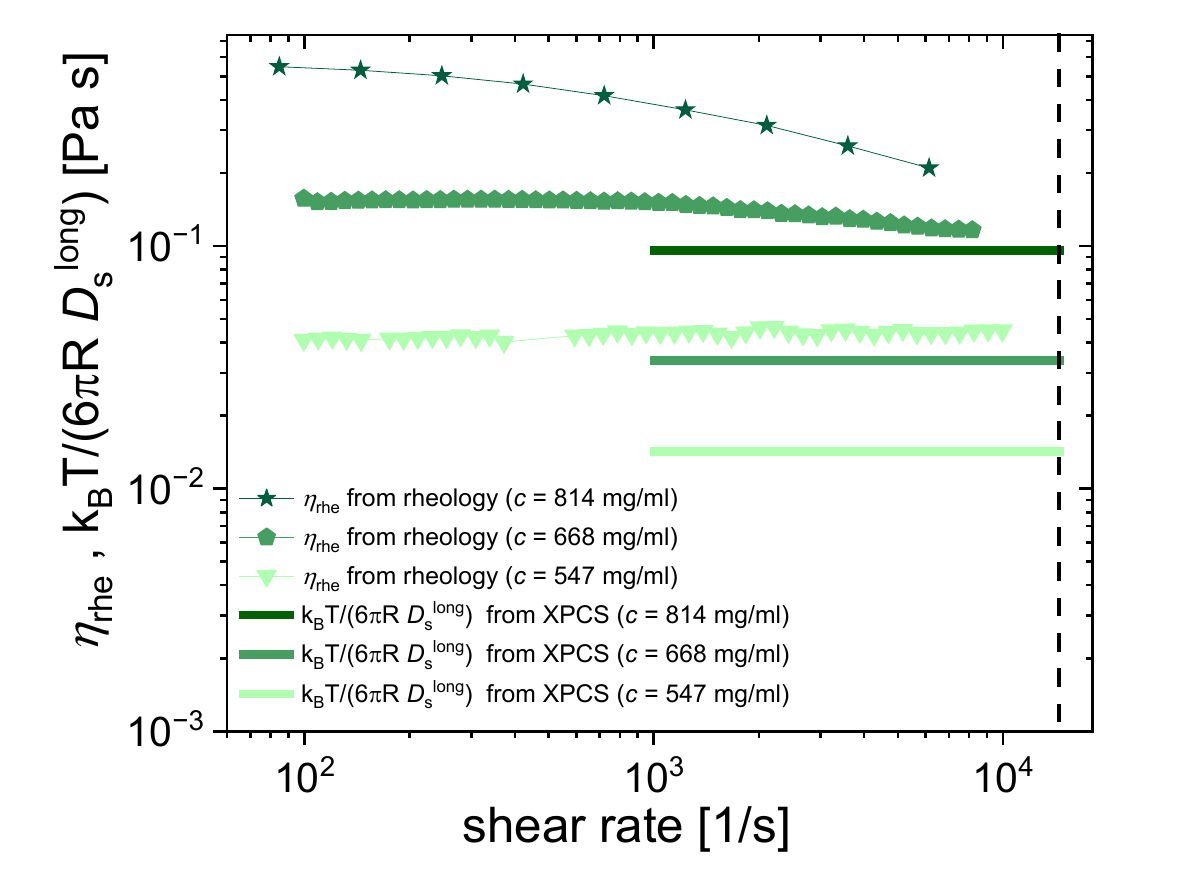} 
	\caption{\textbf{Comparing viscosity from rheology and $k_BT/(6\pi R D_\mathrm{s}^\mathrm{long})$ from XPCS:} The viscosity of yolk-plasma samples at three concentrations measured using rheology, $\eta_\mathrm{rhe}$, compared with $k_BT/(6\pi R D_\mathrm{s}^\mathrm{long})$ values from XPCS. The dashed vertical line indicates the $\omega$ = 1/69 µs$^{-1}$ (the XPCS measurement window is 0.22--69 µs). The long-time linear behavior of MSD is observed close to 69 µs, as depicted in Fig.~\ref{fig:MSD}A, and hence a diffusion coefficient of $D_\mathrm{s}^\mathrm{long}$ is expected for timescales higher than $\approx$ 69 µs.
    }
\label{figSI:viscosity} 
\end{figure}

Shear-dependent viscosity measurements were performed using a Bolin Gemini
rotational HR nano-rheometer. The samples were sheared using a cone plate with an angle of 1$^{o}$ (angle formed between the cone and its base) and a diameter of 40 mm. Due to limited shear ranges of the instrument and spilling of sample at high shear rates, the experiments were limited to shear rates below $\approx 10^{4}$ 1/s. In Fig.~\ref{figSI:viscosity}, the estimated values of viscosity from rheology, $\eta_\mathrm{rhe}$, for all three samples, along with $k_BT/(6\pi R D_\mathrm{s}^\mathrm{long})$ estimated from XPCS, are given.  

\section{Contribution of long-range hydrodynamic interactions to long-time diffusion}

\begin{figure} 
	\centering
	\includegraphics[width=0.8 \textwidth]{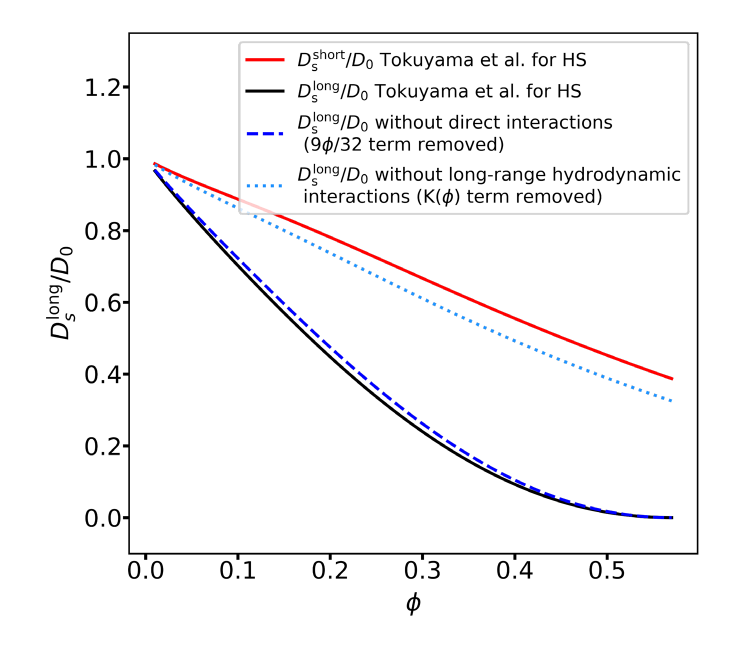} 
	\caption{\textbf{Long-time diffusion: } The contribution of direct and long-range hydrodynamic interactions to the long-time self-diffusion. red curve: Eq. \ref{eq:Ds_D0_tokuyama}, black solid curve: Eq. \ref{eq:DL_D0_tokuyama} with $\epsilon=1$. blue dashed curve: Eq. \ref{eq:DL_D0_tokuyama} without direct interactions. blue dotted curve: Eq. \ref{eq:DL_D0_tokuyama} without $K(\phi)$ term. 
}
\label{figSI:SI_Tokuyama_DL_Do} 
\end{figure}

In Fig.~\ref{figSI:SI_Tokuyama_DL_Do}, the contribution of direct interactions and long-range hydrodynamic interactions to the long-time self-diffusion coefficient is depicted. 

\section{Experimental parameters of SAXS measurements at ESRF}

The SAXS measurements were performed at ID02 beamline at ESRF, and the experimental parameters are given in Table.~\ref{tabSI:esrf}.

\begin{table} 
	\centering
	\caption{\textbf{Details of SAXS experiment parameters at ESRF}}
	\begin{tabular}{|l|c|c|c} 
 \hline
        X-ray energy [keV] & 12.23 \\
		beam size [µm $\times\,$µm] & 40$\times$40  \\
		detector & EIGER2 4M \\
        pixel size [µm $\times\,$µm] & 75$\times$75 \\
        sample to detector distance [m] & 1.5\\
		\hline
	\end{tabular}
\label{tabSI:esrf}
\end{table}

\clearpage

\section{Assessment of X-ray induced beam damage}

In X-ray scattering measurements, exposing the material to X-rays above a critical dosage causes radiation-induced damage. In XPCS experiments, X-ray beam-induced damage is characterized by the changes in the native structure and equilibrium dynamics \cite{ruta2017hard,pintori2019relaxation,marioNatComm}. Recent investigations reveal that beam-induced effects depend on both cumulative dosage and dose rate \cite{marioNatComm}. To eliminate beam-induced artifacts in XPCS measurements, it is necessary to determine the critical dose and critical dose rate thresholds for a specific sample system and execute subsequent experiments below these limits. To determine the critical dose and dose rate of yolk-plasma, we performed XPCS measurements at various incident X-ray intensities such that the dose rate on the sample is different for the same exposure time. The dose absorbed by the sample is estimated using the relation \cite{moller2019x},
\begin{equation}
    \mathrm{Dose} = \frac{\Phi E A_\text{cof}}{d \; b^{2} \; \rho_\mathrm{m}},
\end{equation}
\noindent where $E, \; d, \; b$, and $ \rho_\mathrm{m}$ are the energy of the incident X-ray, the thickness of the capillary, the beam size, and the density of water (see Table.~\ref{tabSI:beamtime}). $A_\text{cof}$ is the attenuation coefficient of the sample. The attenuation coefficient of water was used for the estimation of the dose.  The estimated values of the dose rate corresponding to different values of X-ray transmissions during two different beamtimes can be found in Table.~\ref{tabSI:doserate}.

\begin{table} 
	\centering
	\caption{\textbf{Details of XPCS experiment parameters at EuXFEL}}
	
	\begin{tabular}{|l|c|c|c} 
 \hline
		parameter  & beamtime id & beamtime id \\
                   & 5397 & 6996 \\
		\hline
        X-ray energy [keV] & 10 & 9 \\
		beam size [µm $\times\,$ µm]& 12$\times$12 & 15$\times$15 \\
		pulse energy [µJ]& $\approx$ 1640 & $\approx$ 460 \\
        inter-pulse time [ns] & 222 & 444 \\
		pulses per train & 310 & 280 \\
        sample to detector distance [m]& 7.68 & 7.265\\
		\hline
	\end{tabular}
\label{tabSI:beamtime}
\end{table}

\begin{table} 
	\centering
	\caption{\textbf{Details of samples and physical quantities}}	
	\begin{tabular}{|c|c||c|c|} 
		\hline
		Transmission  & Dose rate [kGy/µs] & Transmission & Dose rate [kGy/µs] \\
                   & beamtime id: 5397 &    & beamtime id: 6996 \\
        \hline
        1.34$\times 10^{-5}$ & 0.23 & 4.638$\times 10^{-5}$ &  0.09\\
              -              & &     9.73$\times 10^{-5}$     & 0.19\\
              -              & &     2$\times 10^{-4}$        & 0.38\\
              -              & &     7.89$\times 10^{-4}$     & 1.46\\

\hline
	\end{tabular}
\label{tabSI:doserate}
\end{table}

\begin{figure} 
	\centering
	\includegraphics[width=0.95 \textwidth]{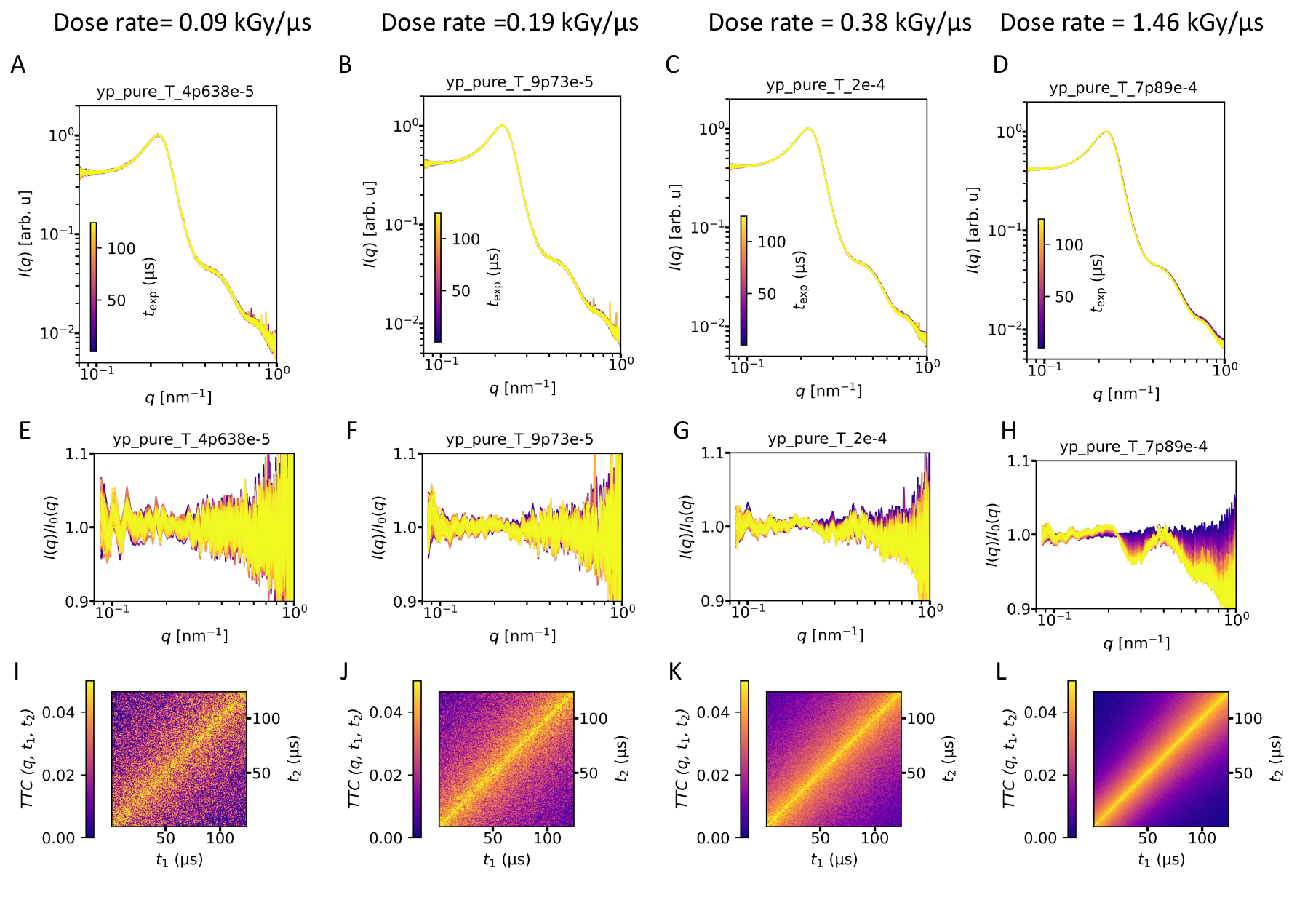} 
	\caption{\textbf{Beam-induced effects on structure and dynamics at different dose rates:} Top row: Temporal evolution of $I(q)$ as a function of $q$ for different dose rate values as mentioned in the figure title. The scale bar represents the sample exposure time in a scan ($t_\text{exp}$). Middle row: normalized $I(q)$ profiles of the top row with respect to the first profile $I_{0} (q)$. The scale bar shown in the top row figures also applies to the corresponding middle row figures. Here, any deviation exceeding the noise-level from $I(q) / I_{0} (q) = 1$ is considered as the change in the structure caused by the beam. With increasing total absorbed dose, the scattering intensity at high-$q$ seems to change, as depicted in \textbf{H}. Bottom row: TTCs at $q = 0.225 \, \text{nm}^{-1}$ extracted for measurements of various dose rates as mentioned in the legend (at the top of the first row).
}
	\label{figSI:beam_induced_TTC_Iq} 
\end{figure}

\begin{figure} 
	\centering
	\includegraphics[width=0.6 \textwidth]{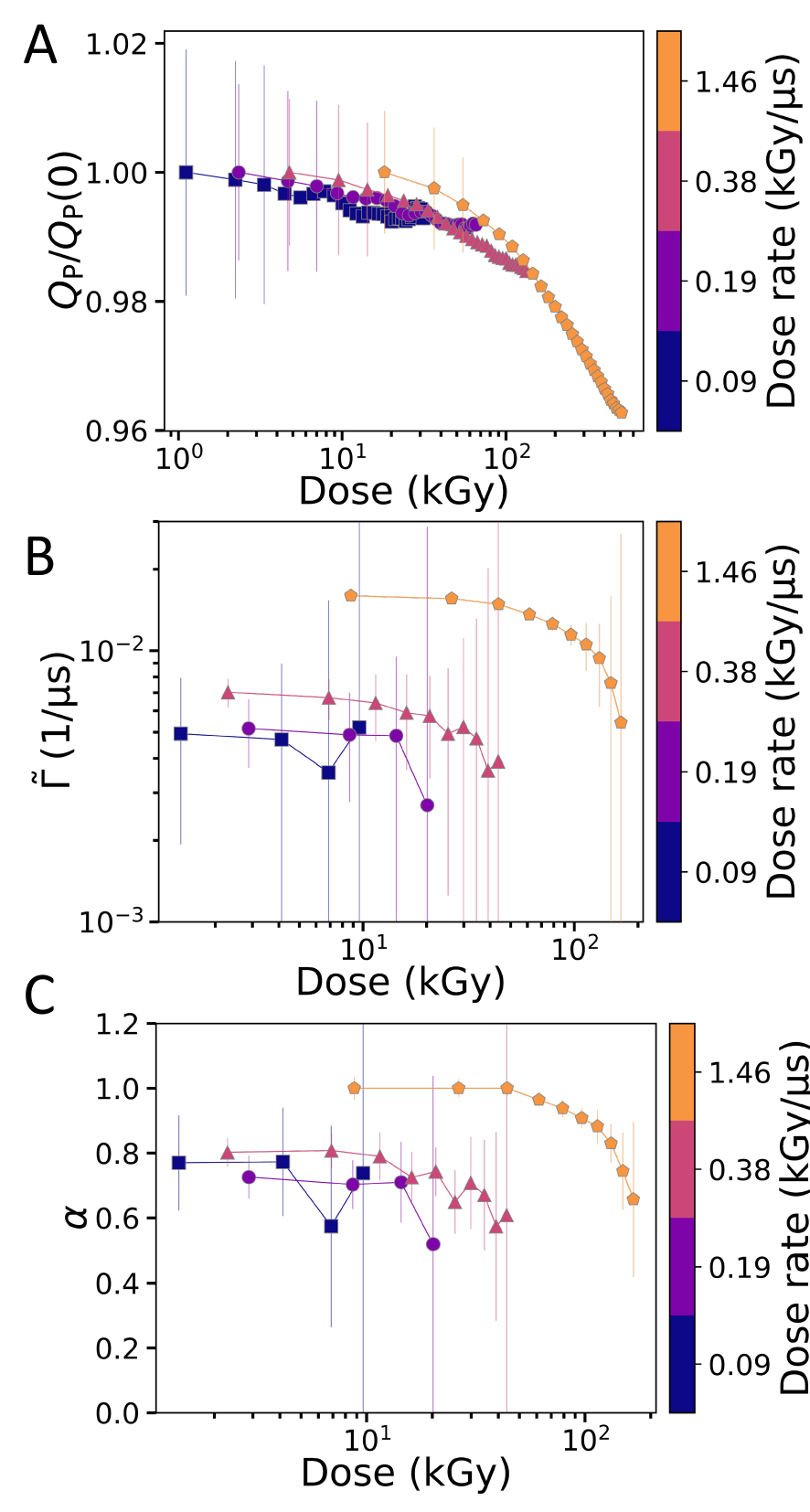} 
	\caption{\textbf{Beam induced effects on structure and dynamics: } (\textbf{A}) The normalized Porod invariant (Eq.~\ref{eqSI:Q_P}) with respect to initial Porod invariant ($Q_\mathrm{P}(0)$) as a function of accumulated dose. 
    The relaxation rate (\textbf{B}) and $\alpha$ (\textbf{C}) at $q = 0.225 \,\text{nm}^{-1}$, as a function of dose for different dose rates as depicted by the color scale. 
}
	\label{figSI:beam_induced_QP_Gamma} 

\end{figure}

The $I(q)$ as a function of X-ray exposure time at different dose rates is provided in Fig.~\ref{figSI:beam_induced_TTC_Iq}A-D. To evaluate further, a normalized $I(q)$ with respect to the first $I(q)$ curve is shown in Fig.~\ref{figSI:beam_induced_TTC_Iq}E-H. At low dose rates, there is almost no change in scattering intensity with increasing total accumulated dose within the experimental time, while at a dose rate of 1.46 kGy/µs, the $I(q)$ shows a continuous increase with longer exposure. To quantify further, we estimate the Porod invariant \cite{marioNatComm} using

\begin{equation}
Q_\text{P} = \int ^{q_{2}} _{q_{1}}  q^{2} \; I(q)\; \text{d}q.
\label{eqSI:Q_P}
\end{equation}
\noindent The lower and upper limit of integration are $q_{1}$ = 0.3 nm$^{-1}$ and  $q_{2}$ = 0.8 nm$^{-1}$ respectively. This range is chosen as the beam-induced structural changes start to be observed in this region, as shown in Fig.~\ref{figSI:beam_induced_TTC_Iq}H. The $Q_\text{P}$ normalized with respect to the initial value is depicted in Fig. \ref{figSI:beam_induced_QP_Gamma}A. Interestingly, the normalized $Q_\text{P}$ decreases after a deposited dose of $\approx$ 20 kGy, thus giving an upper estimate for the critical dose for the egg yolk samples. Next, we estimate the changes in the dynamical parameters of the sample. Representative TTCs extracted for XPCS scans with different dose rates are shown in Fig. \ref{figSI:beam_induced_TTC_Iq}, bottom row. Clearly, the dynamical slowdown is observed at the highest dose rate of 1.46 kGy/µs, as concluded from the broadening of the diagonal region in the $TTC$. To quantify the changes in dynamics, we extracted autocorrelation functions and modeled them using Eq.~\ref{eq:g2}. The extracted $\tilde{\Gamma} (q)$ and KWW values are depicted in Fig. \ref{figSI:beam_induced_QP_Gamma}B and C, respectively. Equilibrium dynamics is observed up to a dose rate of 0.38 kGy/µs and a total dose of 20 kGy. Hence, all measurements shown in the main text were performed below the critical threshold of 20 kGy using a dose rate of 0.23 kGy/µs.

Further, X-ray induced heating effects \cite{dallari2024real, marioNatComm} were evaluated by estimating the temperature increase induced by a pulse with energy $E_\mathrm{p}$ as,
\begin{equation}
    \Delta T = \frac{4\, \mathrm{ln}(2) E_\mathrm{p} (1-\mathrm{e}^{-d/\mu_{0}})}{ 2 \pi c_\mathrm{p} \,\rho_\mathrm{m} \, b^2 d}
\end{equation}
\noindent where $c_\mathrm{p} $ and $\mu_{0}$ are specific heat capacity and effective attenuation length. For the estimation of $\Delta T$, $c_\mathrm{p} $ = 3120 J/kg K (of egg yolk \cite{abbasnezhad2016numerical}) and the attenuation length of water was used. The material density of egg yolk $\rho_\mathrm{m} \approx$ 1 g/cm$^{3}$ (from ref. \cite{rita2007effect}) is used for the calculation. The increase in temperature of the solution after exposure to 310 pulses (a full train) during the experiment (experiment number 5397) is estimated to be $\approx$ 2 K. Such a minor change in temperature is insufficient to cause any protein denaturation and can thus be safely neglected.

\end{document}